\documentclass[10pt,journal,compsoc]{IEEEtran}
\ifCLASSOPTIONcompsoc
  \usepackage[compress]{cite}
\else
  \usepackage{cite}
\fi
\ifCLASSINFOpdf
   \usepackage[pdftex]{graphicx}
\else
\fi
\usepackage{amsmath, bm}
\usepackage{threeparttable}

\usepackage{algorithm}
\usepackage{algorithmic}
\floatname{algorithm}{Alg.}
\DeclareMathOperator*{\argmax}{arg\,max}

\usepackage{color}
\definecolor{turquoise}{cmyk}{0.65,0,0.1,0.1}
\definecolor{purple}{rgb}{0.45,0,0.65}
\definecolor{darkgreen}{rgb}{0.0, 0.5, 0.0}
\definecolor{darkred}{rgb}{0.4, 0.0, 0.0}
\definecolor{darkblue}{rgb}{0.0, 0.0, 0.4}
\definecolor{blue}{rgb}{0.0, 0.0, 0.8}

\newcommand{\ignore}[1]{}

\newcommand{\hide}[1]{{}}

\usepackage{picins}
\usepackage[normalem]{ulem}
\begin{document}

\title{Error-Bounded and Feature Preserving Surface Remeshing with Minimal Angle Improvement}

\author{Kaimo~Hu,
        Dong-Ming~Yan,
        David Bommes,
        Pierre Alliez
        and~Bedrich~Benes % <-this % stops a space
\IEEEcompsocitemizethanks{
\IEEEcompsocthanksitem K. Hu is with Purdue University, 610 Purdue Mall, West Lafayette, IN 47907, US. E-mail: hukaimo02@gmail.com.
\IEEEcompsocthanksitem D.-M. Yan is with the National Laboratory of Pattern
Recognition, Institute of Automation, Chinese Academy of Sciences, Beijing
100190, China. E-mail: yandongming@gmail.com.
\IEEEcompsocthanksitem D. Bommes is with RWTH Aachen University, Schinkelstr. 2, 52062 Aachen, Germany. E-mail: bommes@aices.rwth-aachen.de.
\IEEEcompsocthanksitem P. Alliez is with Inria Sophia-Antipolis – Mediterranee, 2004 route des Lucioles BP 93, 06902 Sophia-Antipolis cedex, France. E-mail: pierre.alliez@inria.fr.
\IEEEcompsocthanksitem B. Benes is with Purdue University, 610 Purdue Mall, West Lafayette, IN 47907, US. E-mail: bbenes@purdue.edu.}
}

\IEEEtitleabstractindextext{%
\begin{abstract}
The typical goal of surface remeshing consists in finding a mesh that is (1) geometrically faithful to the original geometry, (2) as coarse as possible to obtain a low-complexity representation and (3) free of bad elements that would hamper the desired application.
%(e.g.~the minimum interior angle is above an application-dependent threshold). Unfortunately, these three optimization goals are in conflict with each other such that they cannot be easily optimized simultaneously. Consequently, in one way or another all remeshing algorithms must prioritize these goals. 
In this paper, we design an algorithm to address all three optimization goals simultaneously. The user specifies desired bounds on approximation error $\delta$, minimal interior angle $\theta$ and maximum mesh complexity $N$ (number of vertices). Since such a desired mesh might not even exist, our optimization framework treats only the approximation error bound $\delta$ as a hard constraint and the other two criteria as optimization goals. More specifically, we iteratively perform carefully prioritized local operators, whenever they do not violate the approximation error bound and improve the mesh otherwise. Our optimization framework greedily searches for the coarsest mesh with minimal interior angle above $\theta$ and approximation error bounded by $\delta$.
%Since large angle bounds $\theta$ sometimes require excessive mesh refinement or are even infeasible, refinement operators are not performed if the complexity is already beyond the user specified limit $N$. 
Fast runtime is enabled by a local approximation error estimation, while implicit feature preservation is obtained by specifically designed vertex relocation operators. 
Experiments show that our approach delivers high-quality meshes with implicitly preserved features and better balances between geometric fidelity, mesh complexity and element quality than the state-of-the-art.
\end{abstract}

}

\maketitle

% is not selected <OR> if conference mode is selected - because compsoc
% conference papers position the abstract like regular (non-compsoc)
\IEEEdisplaynontitleabstractindextext

% \fi
\IEEEpeerreviewmaketitle

\ifCLASSOPTIONcompsoc
\IEEEraisesectionheading{\section{Introduction}\label{sec:introduction}}
\else

\section{Introduction}
\label{sec:introduction}
\fi

\begin{figure*}[!ht]
    \centering
    \begin{minipage}{25.2mm}
        \centering
        \includegraphics[width=25.1mm]{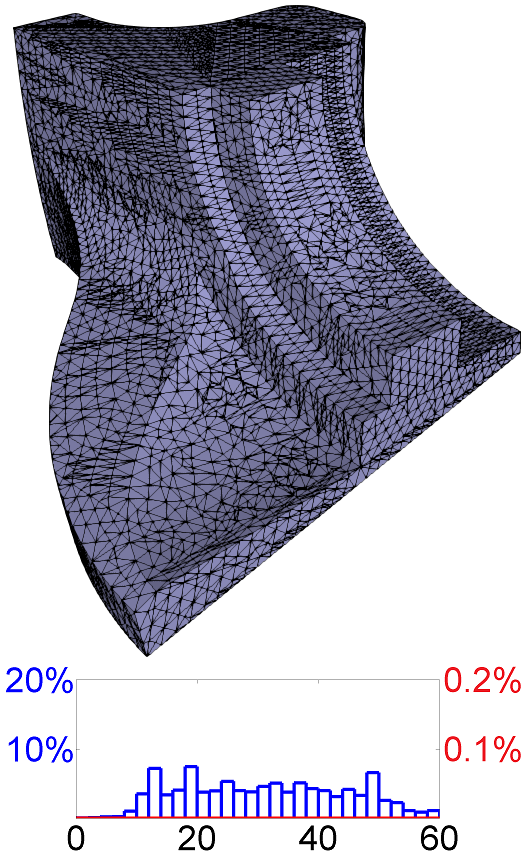}\\
    {\small (a) Input model.}
    \end{minipage}
    \begin{minipage}{25.2mm}
        \centering
        \includegraphics[width=25.1mm]{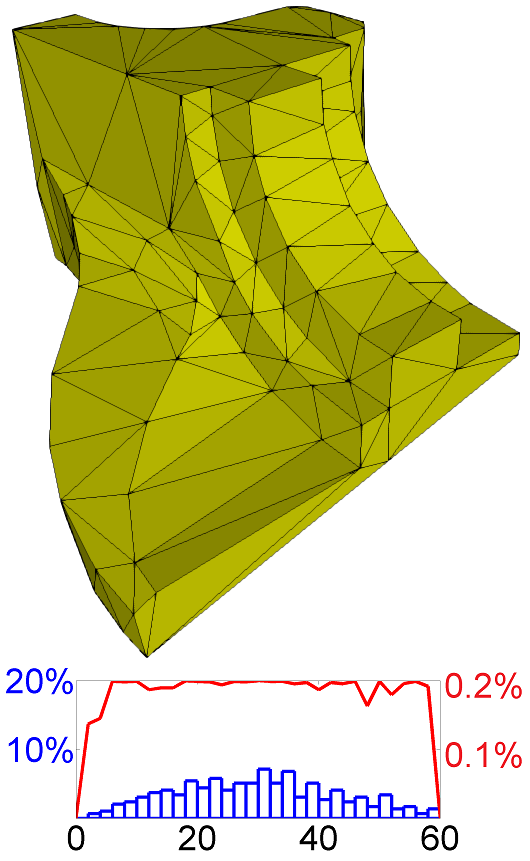}\\
    {\small (b) $\theta = 0^\circ$.}
    \end{minipage}
    \begin{minipage}{25.2mm}
        \centering
        \includegraphics[width=25.1mm]{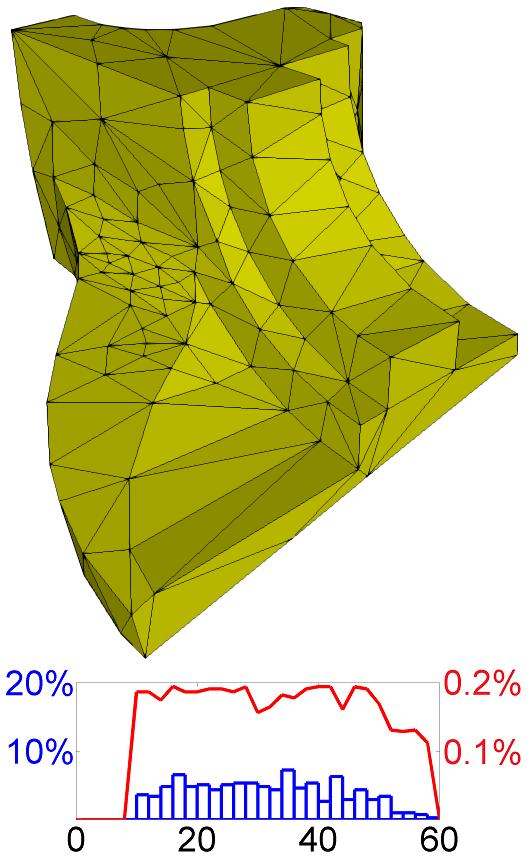}\\
    {\small (c) $\theta = 10^\circ$.}
    \end{minipage}
    \begin{minipage}{25.2mm}
        \centering
        \includegraphics[width=25.1mm]{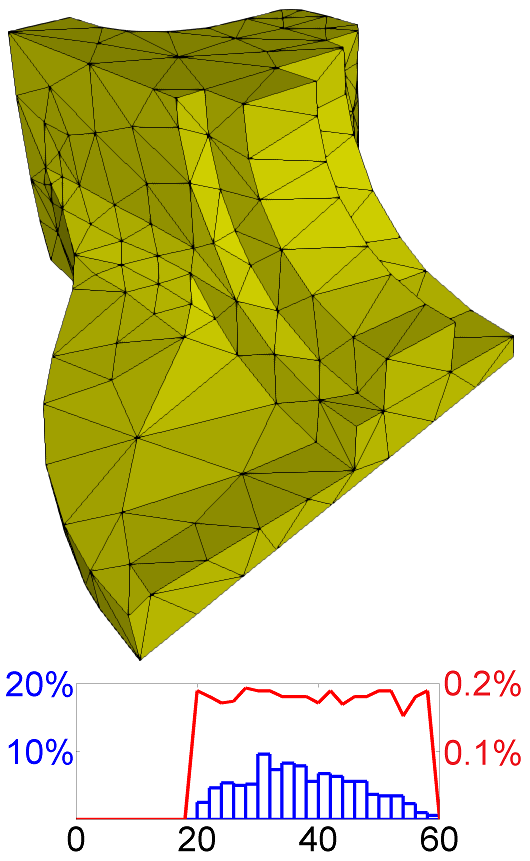}\\
    {\small (d) $\theta = 20^\circ$.}
    \end{minipage}
    \begin{minipage}{25.2mm}
        \centering
        \includegraphics[width=25.1mm]{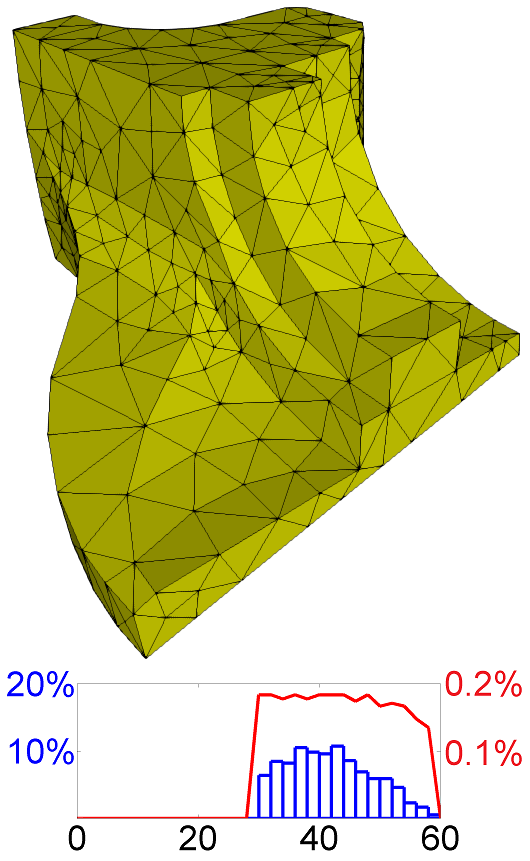}\\
    {\small (e) $\theta = 30^\circ$.}
    \end{minipage}
    \begin{minipage}{25.2mm}
        \centering
        \includegraphics[width=25.1mm]{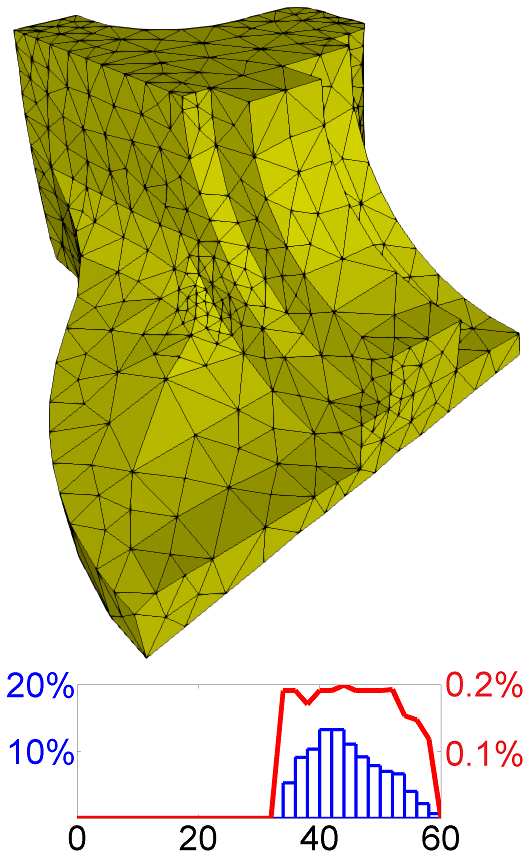}\\
    {\small (f) $\theta = 35^\circ$.}
    \end{minipage}
    \begin{minipage}{25.2mm}
        \centering
        \includegraphics[width=25.1mm]{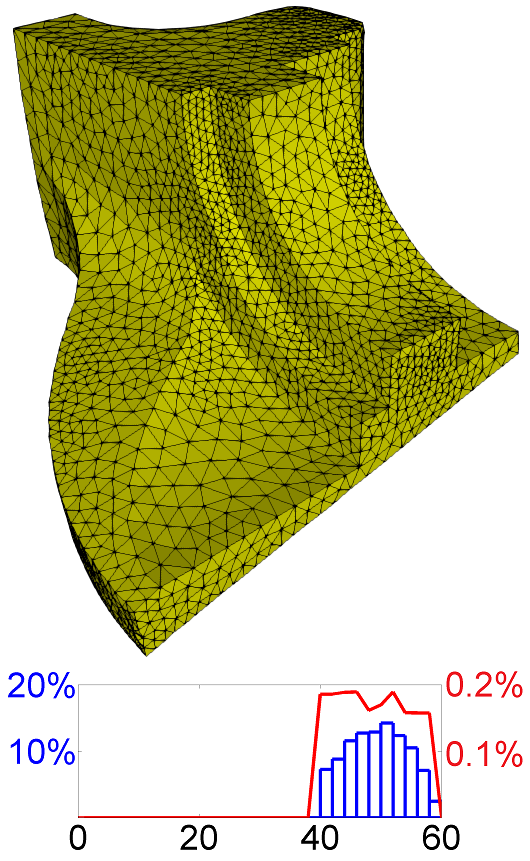}\\
    {\small (g) $\theta = 40^\circ$.}
    \end{minipage}\\
    %\vskip0.1in
    \caption{Examples of surface meshes generated with our approach. The input model (a) has 7.2k vertices. From (b) to (g) are the results with different minimal angle threshold $\theta$. The angle distributions and their corresponding approximation error are shown below each result, in blue bars and red curves respectively. The error-bound threshold~$\delta$ is set to $0.2\%$ of the diagonal length of the input's bounding box $(\%bb)$.}
\label{fig:varying_theta}
\end{figure*}

\IEEEPARstart{S}urface remeshing is a key component in many geometry processing applications such as simulation, deformation, or parametrization~\cite{Botsch10}. While many remeshing techniques are goal-specified, a common goal is to find a satisfactory balance between the following three criteria:

\begin{itemize}
\item The output mesh should be a good approximation of the input, making the \textbf{geometric fidelity}, usually measured as the approximation error, a key requirement for most applications.

\item The \textbf{quality of mesh elements} is crucial for robust geometry processing and numerical stability of simulations, which requires fairly regular meshes in terms of both geometry and connectivity. Particularly, a lower bound on the minimal angle is vital for many simulation applications~\cite{Shewchuk02}.

\item\textbf{Mesh complexity}, measured as the number of mesh elements, is important for an efficient representation of complex shapes. Since mesh complexity usually conflicts with geometric fidelity and element quality, a ``just enough'' resolution for the required element quality and geometric fidelity should be the goal.
\end{itemize}

However, to the best of our knowledge, only a few approaches fulfill all of the above criteria. Most existing methods that generate meshes with high element quality often require high mesh complexity or introduce high approximation error~\cite{Alliez05, Valette08}. The error-driven methods, while preserving the results in controllable geometric fidelity and low mesh complexity, do not deal with the element quality~\cite{Cohen04,Mandad15}. In addition, many approaches require the sharp features to be specified or detected in advance~\cite{Alliez02,Yan09,Yan13}, which is usually difficult and error-prone.\\

We propose an approach that controls both the approximation error and element quality simultaneously. Our approach only requires the user to specify the error-bound threshold $\delta$, the desired minimal angle $\theta$ and an upper bound on the mesh complexity $N$, measured as the number of vertices. In an initial phase our algorithm coarsens the mesh as much as possible while respecting the error-bound $\delta$.
It then iteratively improves the minimal angle of the mesh until the desired bound for $\theta$ or $N$ is met.  
Since all atomic operations respect the error-bound $\delta$, the result is guaranteed to satisfy geometric fidelity. In contrast, low mesh complexity and high quality of mesh elements are optimization goals, which depending on the model and the desired bounds might or might not be met. However, experiments show that given a reasonable vertex budget $N$ and an angle bound $\theta \leq 35^\circ$ our algorithm is usually able to reach all three goals simultaneously.
Moreover, our method preserves geometric features like sharp creases and ridges without explicitly specifying or detecting them.

Our method is inspired by the remeshing methods that proceed by applying a series of local operators such as edge split, edge collapse, edge flip, and vertex relocate~\cite{Hoppe93,Botsch04}. Contrary to the existing methods that iteratively apply these operations sequentially and globally, our core algorithm employs a dynamic priority queue to maintain all angles which are smaller than the user specified threshold, and then greedily applies local operators whenever an improvement of the mesh is possible. After the initial coarsening our method only improves the mesh in local regions where the element quality is poor, and thus modifies the model as little as necessary with respect to the minimal angle threshold $\theta$ and the error-bound constraint $\delta$.

Since we apply local operators directly on the mesh, without using any surface parameterizations or density functions, our algorithm is naturally suitable for high genus, irregular, and multi-component inputs, as well as meshes with very high/low resolutions. In summary, we claim the following contributions:
\begin{itemize}
  \item A surface remeshing algorithm with minimal angle improvement based on applying local operators, which bounds the ``worst element quality'' (Sec.~\ref{sec:algorithm_overview}).
  \item A reliable and efficient local error update scheme based on approximated symmetric Hausdorff distance, which bounds the geometric fidelity (Sec.~\ref{sec:local_error}).
  \item Two feature intensity functions designed for vertex position relocation, which enable implicit feature preservation and support geometric fidelity (Sec.~\ref{sec:feature_function_based_vertex_position}).
\end{itemize}

\section{Related Work}
\label{sec:related_work}

The variety of applications leads to a large number of different remeshing techniques. We restrict the discussion to the most relevant aspects of our algorithm, i.e.~high-quality remeshing, error-driven remeshing and feature-preserving remeshing. For a more complete discussion we refer the reader to the survey~\cite{Alliez08}.

\noindent\textbf{High quality remeshing} is typically based on sampling and Centroidal Voronoi Tessellation (CVT) optimization~\cite{Du99}. Early approaches apply 2D CVT in a parametric domain~\cite{Alliez03A, Alliez05, Alliez02, Surazhsky03A, Surazhsky03B, Fuhrmann10}. Instead of CVT optimization, Vorsatz et al.~\cite{Vorsatz03} utilize a partial system approach in the parametric domain. In general, parametrization-based methods suffer from the additional distortion of the map and the need to stitch parameterized charts for high genus surfaces. Valette et al.~\cite{Valette08} perform a discrete version of CVT directly on the input surface. However, the resulting mesh quality can be poor due to the inexact computation. Yan et al.~\cite{Yan09, Yan14A, Yan16} avoid the parameterization by computing the 3D CVT restricted to the surface. Additionally, they proposed blue-noise remeshing techniques using adaptive maximal Poisson-disk sampling~\cite{Yan13,Guo15}, farthest point optimization~\cite{Yan14B}, and push-pull operations~\cite{Ahmed16}, which improve the element quality as well as introducing blue-noise properties. However, these approaches still suffer from common limitations, e.g., geometric fidelity and the minimal angle cannot be explicitly bounded. Moreover, sharp features must be specified in advance.

Another way to avoid the stitching problem is to operate directly on the surface mesh~\cite{Frey00, Narain12}. An efficient isotropic approach proposed by Botsch and Kobbelt \cite{Botsch04} takes an edge length as input and repeatedly splits long edges, collapses short edges, equalizes vertex valences and relocates vertex positions until all edges are approximately of the specified target edge length. To extend this work to an adaptive version, Dunyach et al.~\cite{Dunyach13} replace the constant target edge length with an adaptive sizing field that is sensitive to local curvatures. Since this kind of methods requires neither surface parameterization nor density functions, they are easy to implement, robust for high genus inputs, and efficient for real-time applications. Our method falls into this category. However, we enrich the local operators and apply them in a more selective manner in order to obtain guarantees on the geometric fidelity as well as higher-quality results and implicit feature preservation.

\noindent\textbf{Error-driven remeshing} amounts to generating a mesh that optimizes the tradeoff between geometric fidelity and mesh complexity. Cohen-Steiner et al.~\cite{Cohen04} propose an error-driven clustering method to coarsen the input mesh. They formulate the approximation problem as a variational geometric partitioning problem, and optimize a set of planes iteratively using Lloyd's iteration~\cite{LLoyd82} to minimize a predefined approximation error.

The mesh simplification techniques are similar to error-driven remeshing in some way. Garland and Heckbert~\cite{Garland97} use iterative contractions of vertex pairs to simplify models and maintain surface approximation error based on quadric error metrics. Borouchaki and Frey~\cite{Borouchaki05} define a fidelity metric named Hausdorff envelope, and simplify and optimize the reference mesh that stays inside the tolerance volume. While they consider the geometric fidelity and element quality simultaneously, nothing is done to improve the worst element quality. Our method guarantees the worst element quality by improving the minimal angle and keeping the mesh complexity as low as possible, with respect to a given error-bound.
Based on the concept of tolerance volume, Mandad et al.~\cite{Mandad15} propose an isotopic approximation method. Their algorithm generates a surface triangle mesh guaranteed to be within a given tolerance volume. However, they generate the mesh from scratch, and mainly focuse on mesh complexity rather than element quality. Instead, we strive for good balances between mesh complexity and element quality with a given input.

\noindent\textbf{Feature preservation} is crucial in remeshing. However, automatically identifying sharp features on a surface mesh is a difficult problem that depends both on the local shape and on the global context and semantic information of the model. This makes feature detection a strongly ill-posed problem. A wide range of approaches address this problem~\cite{sun02,jiao02,Chen13}. However, none of them works reliably for all kinds of meshes. Most remeshing algorithms avoid this problem by assuming that the features have been specified in advance~\cite{Yan09, Yan13, Yan14A, Yan14B, Zhong13, Guo15}. Some remeshing techniques try to preserve features implicitly~\cite{Kobbelt03,Levy10}. Vorsatz et al.~\cite{Vorsatz01} first apply a relaxation in the parameter domain, and then snap the vertices to feature edges and corners. Since they separate remeshing and feature-snapping, the resulting mesh quality near sharp features might be poor. Valette~\cite{Valette08} alleviates this issue by embedding the Quadric Error Metric approximation (QEM) criterion inside the CVT optimization. However, the performance of their cluster-based method is highly dependent on the quality of the input, and the sharp features might not be well preserved when users specify a small vertex budget. Jakob et al.~\cite{Jakob15} propose a general framework for isotropic triangular/quad-dominant remeshing using a unified local smoothing operator, in which the edges naturally align to sharp features. However, little attention is paid on the approximation error and element quality. We address this problem by optimizing the element quality explicitly in combination with implicit feature preservation based on the new defined feature intensity functions.

\section{Algorithm Overview}
\label{sec:algorithm_overview}

Given a 2-manifold triangular mesh $M_I$, the goal of our algorithm consists in finding an improved surface mesh $M_R$ with approximation error below $\delta$, minimal interior angle above $\theta$ and mesh complexity below $N$. The main idea is to transform the mesh by a series of discrete local operators as illustrated below:

\begin{figure}[hbt]
\centering
\includegraphics[width=0.95\linewidth]{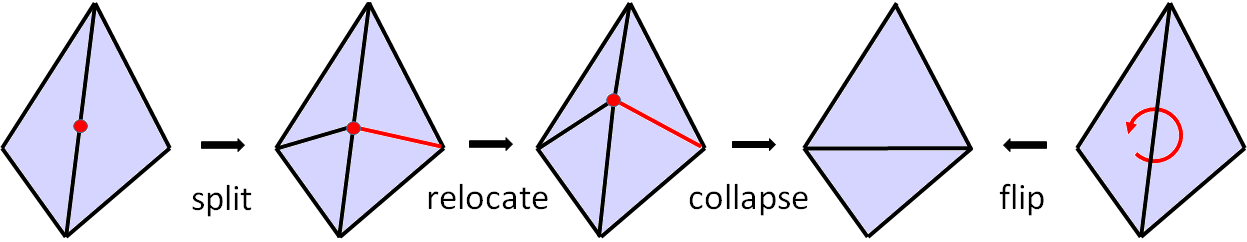}
\caption{Common local operators~\cite{Botsch10}.}
\label{fig:flip}
\end{figure}

In our algorithm, only \textbf{edge collapse}, \textbf{edge split} and \textbf{vertex relocation} operators are employed. Edge flips are implicitly performed as a combination of an edge split followed by an edge collapse (cf.~Fig.~\ref{fig:flip}). This convention not only lowers the combinatorial complexity of all operators but is also advantageous for our implicit feature preservation since edge flips tend to destroy sharp creases and would require additional nontrivial checks.

The remeshing algorithm is designed to perform local operators in a conservative manner. More specifically, a local operator is only executed if it respects the approximation error bound $\delta$, does not introduce new interior angles below current minimal angle $\theta_{min}$ and maintains the 2-manifoldness of the mesh. This behavior can be interpreted as a conservative greedy algorithm, which in each step identifies the most promising operation that improves the result (either coarsens or improves angles), while never leaving the feasible set of meshes with approximation error below $\delta$. Note that since pure topological edge collapses and edge splits improve the mesh quality very rarely, these operations are always combined with vertex position optimization.
%\TODO{discuss that operator are always combined with relocation}

The most crucial design choices of the algorithm are the scheduling of different operators and the efficient modeling and handling of approximation error queries (cf.~Sec.\ref{sec:local_error}). The algorithm passes through three different stages:

The initial phase concentrates on coarsening and runs a mesh simplification, which solely performs edge collapses. 

\begin{figure*}[!ht]
    \centering
    \begin{minipage}{60mm}
        \centering
        \includegraphics[width=55mm]{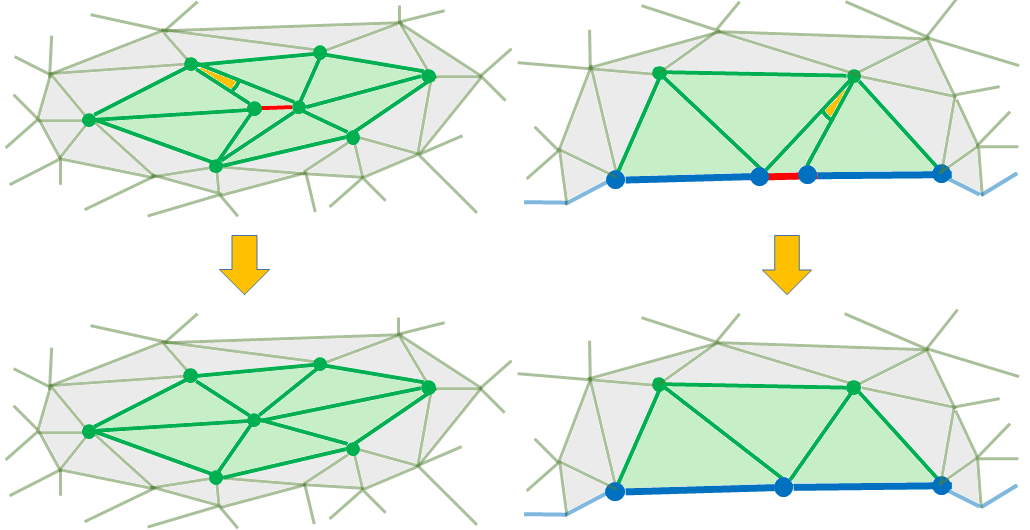}\\
    {\small (a) Edge collapse.}
    \end{minipage}
    \begin{minipage}{60mm}
        \centering
        \includegraphics[width=55mm]{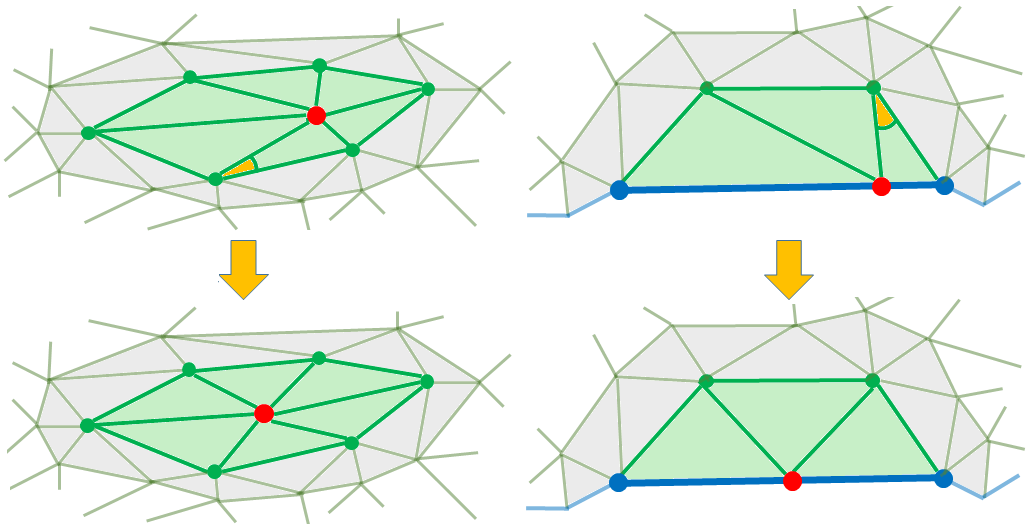}\\
    {\small (b) Vertex relocation.}
    \end{minipage}
    \begin{minipage}{60mm}
        \centering
        \includegraphics[width=55mm]{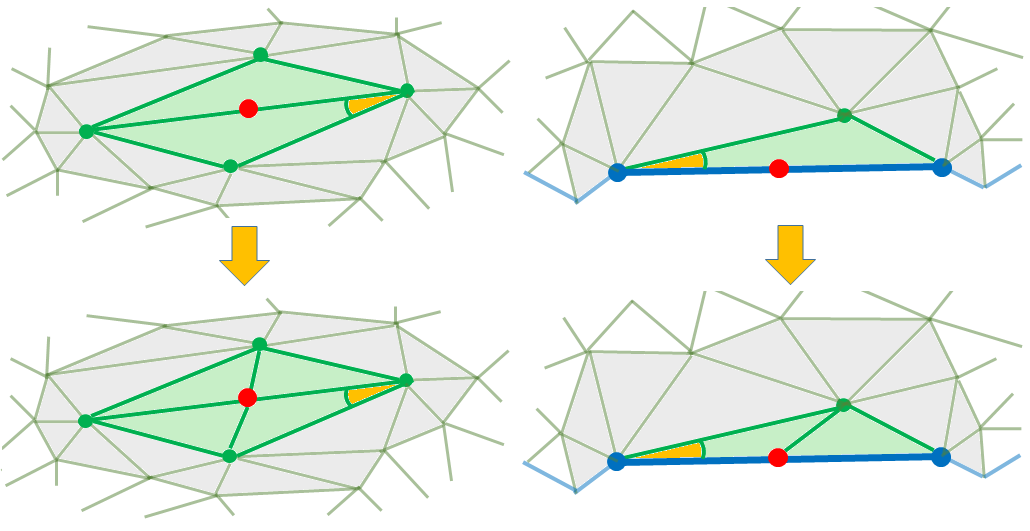}\\
    {\small (c) Edge split.}
    \end{minipage}\\
    \caption{Local operators. The local patches before/after applying the operators are shown in the first/second rows, respectively. The inner patch (green) contains the facets that will be directly affected by the local operators, and the outer patch (gray) includes the facets who share vertices with facets of the inner patch. In each sub figure, the left shows the inner case while the right shows the boundary case, in which the boundary edges and vertices are depicted in blue. We depict the current minimal angle $\theta_{min}$ in yellow.}
\label{fig:operators}
\end{figure*}

The second phase then tries to lift the minimal interior angle above the user-provided bound $\theta$. For this task we devised three different processes, which are tried subsequently, as illustrated in Fig.~\ref{fig:operators}. Since an edge collapse reduces the complexity, it would be the best way to improve the minimal angle. If edge collapse is not possible, we try to improve the minimal angle by relocating one of the triangle's vertices. If vertex relocation also fails, edge splits are considered as the last option as they increase the mesh complexity. While edge splits do not directly improve the minimal angle, they are crucial to enrich the local mesh connectivity in order to enable improvements in subsequent steps. We apply an approach similar to the longest-side propagation path~\cite{Rivara96}, described in more detail in Sec.~\ref{sec:greedy_angle_improvement}.

The third stage of our algorithm freezes the mesh connectivity and concentrates on global vertex relocation. This stage is designed to improve the \emph{average} quality of the mesh elements, while not violating the user-specified bounds. The pseudocode of our remeshing algorithm is shown in Alg.~\ref{alg:remeshing}.

\begin{algorithm}
\caption{$Remeshing(M_I, M_R, \delta, \theta, N)$}
\label{alg:remeshing}
\begin{algorithmic}[1]
\REQUIRE 
$M_I$							$\quad\quad\quad\qquad$\COMMENT{Input mesh}\\
$\delta > 0$					$\quad\quad\quad\qquad$\COMMENT{Error-bound threshold}\\	
$\theta \geq 0$					$\quad\quad\quad\qquad$\COMMENT{Minimal angle threshold}\\
$N > 0$							$\quad\quad\quad\quad$\COMMENT{Desired mesh complexity}\\	
\ENSURE $M_R$					$\quad\quad\quad$\COMMENT{Remeshing result}\\	
\STATE $M_R \leftarrow M_I$;
\STATE $InitialMeshSimplification(M_I, M_R, \delta)$;
\STATE fill $Q$ with angles of $M_R$ smaller than $\theta$;$\quad$\COMMENT{$Q$ is a dynamic priority queue}\\
\STATE $\#V\leftarrow$ the number of vertices in $M_R$;
%\FORALL{halfedge $h \in M_R$}				
%\IF{$\theta_{h} < \theta$}
%\STATE $Q \leftarrow h$;		$\qquad \qquad$ \COMMENT{$\theta_h$ is the opposite angle of $h$}
%\ENDIF
%\ENDFOR
\WHILE{$Q \neq \emptyset \quad {\bf and} \quad \#V < N$}
\STATE $\theta_{min}\leftarrow Pop(Q)$;
\STATE $GreedyImproveAngle(\theta_{min}, M_I, M_R, \delta, \theta)$;
\STATE update $Q$ and $\#V$;
\ENDWHILE
\STATE $FinalVertexRelocation(M_I, M_R, \delta, \theta)$;
\end{algorithmic}
\end{algorithm}
In the following we provide more details on the three phases of initial mesh simplification, greedy angle improvement and vertex relocation.

%--------------------------------------------------------------------
\subsection{Initial Mesh Simplification}
\label{sec:initial_mesh_simplification}

The input mesh is often densely sampled such that the mesh complexity can be significantly reduced without violating the approximation error bound. While such a mesh simplification could potentially also be done on the fly, a specialized pre-process turns out to be more efficient. The goal of this step consists in finding a mesh, which is significantly coarser 
and offers a good starting point for the second phase of element quality improvement.
Simplification is achieved through iteratively collapsing edges and relocating related vertices as long as the approximation error does not exceed $\delta$. Short edges, or those opposite to small angles are likely to lower the mesh quality and are consequently collapsed first. This is achieved through using a priority queue sorted by increasing values of the following function: $l(h) \cdot (\theta_1(h)+\theta_2(h))/2$, where $l(h)$ denotes the length of halfedge $h$ and $\theta_i(h)$ denote the two angles opposite to $h$ and $h$'s opposite halfedge. The pseudocode of the corresponding function is in Alg.~\ref{alg:reducing_complexity}.

\begin{algorithm}
\caption{$InitialMeshSimplification(M_I, M_R, \delta)$}
\label{alg:reducing_complexity}
\begin{algorithmic}[1]
\STATE fill $Q$ with all halfedges;\COMMENT{$Q$ is a dynamic priority queue}
\WHILE{$Q \neq \emptyset$}
\STATE $h\leftarrow Pop(Q)$; $\quad$
\IF{{$CollapseAndRelocateImproves(h, \delta, M_I, M_R)$} }
\STATE {$CollapseAndRelocate(h, \delta, M_I, M_R)$;}
\ENDIF
\STATE update $Q$;
\ENDWHILE
\end{algorithmic}
\end{algorithm}

%--------------------------------------------------------------------
\subsection{Greedy Improvement of Angles}
\label{sec:greedy_angle_improvement}

The second phase is designed to improve the mesh quality by iteratively increasing the smallest angle in the mesh, until the desired angle bound $\theta$ is satisfied or the complexity limit $N$ is reached. Our approach repeats the following process: We simulate a potential operation, test whether the resulting mesh improves and only in this case perform the candidate operation. The mesh improvement test additionally contains several important validity checks.\\

\begin{figure}[hbt]
    \centering
    \begin{minipage}{44mm}
        \centering
        \includegraphics[width=43mm]{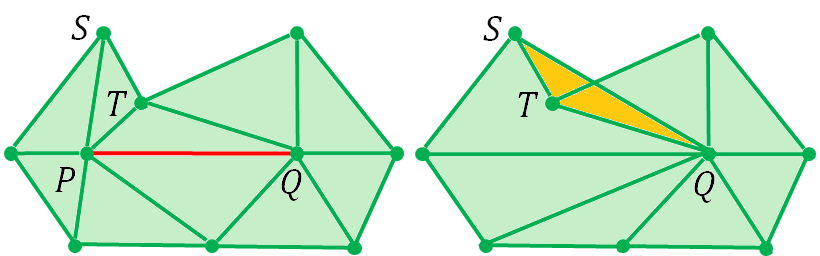}\\
    {\small (a) Edge collapse.}
    \end{minipage}
    \begin{minipage}{44mm}
        \centering
        \includegraphics[width=43mm]{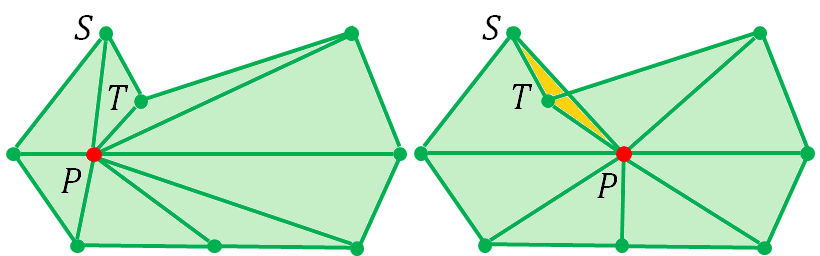}\\
    {\small (b) Vertex relocate.}
    \end{minipage}\\
    \caption{Creation of fold-overs in 2D. (a) Edge $PQ$ will be collapsed into $Q$. $\triangle STQ$ creates fold-overs if $P$ and $Q$ lie on the two sides of line $ST$; (b) Vertex $P$ will be relocated. $\triangle STP$ creates fold-overs if $P$ before and $P$ after relocation are located on the two sides of line $ST$. In 3D, we say that a triangle creates fold-overs if its normals before and after applying a local operator have opposite orientations.}  
\label{fig:facet_flip}
\end{figure}

\noindent\textbf{Mesh Improvement Test}: we simulate each potential operation and measure if the following constraints are satisfied:
\begin{itemize}
  \item \textit{Topology}. For edge collapses topology changes are prevented by checking the \textit{link condition}~\cite{Edelsbrunner06}.  
  \item \textit{Geometry}. The operator should not create fold-overs by flipping the orientation facets (cf. Fig.~\ref{fig:facet_flip}).
  \item \textit{Fidelity}. The approximation error between $M_R$ and $M_I$ should remain below $\delta$ (cf.~Fig.~\ref{fig:stratify_sample}).
\end{itemize}

In order to improve the minimal angle, we greedily apply operators that pass the mesh improvement test. The operators are tested in the following order: edge collapses, vertex relocations and edge splits. The pseudocode of the greedy improvement is shown in Alg.~\ref{alg:greedy_improve}. 

\begin{algorithm}
\caption{$GreedyImproveAngle(\theta_{min}, M_I, M_R, \delta, \theta)$}
\label{alg:greedy_improve}
\begin{algorithmic}[1]
\REQUIRE 
$\theta_{min}$						$\qquad \qquad $ \COMMENT{Angle requiring improvement}\\
$M_I$							$\qquad \qquad \quad$ \COMMENT{Input mesh}\\
$\delta > 0$						$\qquad \qquad \quad$ \COMMENT{Error-bound threshold}\\	
$\theta \geq 0$						$\qquad \qquad \quad$ \COMMENT{Minimal angle threshold}\\
\ENSURE $M_R$					$\qquad \quad$ \COMMENT{Locally improved mesh}\\
\STATE $h\leftarrow$ halfedge opposite to $\theta_{min}$;
\IF{$CollapseAndRelocateImproves(h, \theta_{min}, \theta, \delta, M_I, M_R)$}
\STATE $CollapseAndRelocate(h, \theta_{min}, \theta, \delta, M_I, M_R)$;
\STATE return;
\ENDIF
\STATE let $v_o, v_s$ and $v_e$ be $h$'s opposite, start and end vertex;
\FOR{$v \in \{v_o, v_s, v_e\}$}
\IF{$RelocateImproves(v, \theta_{min}, \theta, \delta, M_I, M_R)$}
\STATE $Relocate(v, \theta_{min}, \theta, \delta, M_I, M_R)$;
\STATE return;
\ENDIF
\ENDFOR
\STATE $h_l \leftarrow LongestSidePropagation(h)$;
\IF{$SplitAndRelocateIsValid(h_l, \theta, \delta, M_I, M_R)$}
\STATE $SplitAndRelocate(h_l, \theta, \delta, M_I, M_R)$;
\ENDIF
\end{algorithmic}
\end{algorithm}

\noindent\textbf{Longest-Side Propagation Path:}
If neither edge collapse nor vertex relocation is possible, we search the longest edge $h_l$ in the neighborhood~\cite{Rivara96} and then split it. Starting from an edge $e$, we iteratively move on to the longest edge of the neighboring two triangles until no further enlargement is possible or we hit the boundary.
%The function in Line 13, Alg.~\ref{alg:greedy_improve} is implemented as follows: initialize $h_l$ as the longer edge incident to $\theta_{min}$, we update it as: if $h_l$ is a boundary edge or the longest edge of its two incident facets, then the update terminates; otherwise, $h_l$ is updated as the longest edge of its two incident facets. 
%This update continues until one of the two termination conditions satisfies. 
Though the edge split operator does not increase $\theta_{min}$ directly, it modifies the local connections such that $\theta_{min}$ can be improved in later iterations (cf. Fig.~\ref{fig:operators}(c)). While there might exist other ways for local connectivity modification, experiments show that this strategy works well in our algorithm.

%--------------------------------------------------------------------
\subsection{Final vertex relocation}
\label{sec:final_vertex_relocation}

To achieve a better overall element quality, in the third phase we perform a series of vertex relocations until the angles can no longer be significantly improved.
In contrast to the second phase which only focuses on specific regions where angle improvement is required, the third stage optimizes all vertex locations.
We maintain the relocation candidate set in a queue, which is initialized with all vertices. Whenever a vertex is relocated, all its neighbors are added to the queue, since a change in the neighborhood might enable further improvements. The pseudocode of the corresponding function is shown in Alg.~\ref{alg:improving_quality}.

\begin{algorithm}
\caption{$FinalVertexRelocation(M_I, M_R, \delta, \theta)$}
\label{alg:improving_quality}
\begin{algorithmic}[1]
\STATE fill $Q$ with all vertices of $M_R$;\COMMENT{$Q$ is a FIFO queue}
\WHILE{$Q \neq \emptyset$}
\STATE $v\leftarrow Pop(Q)$; $\quad$
\IF{{$RelocateImproves(v, \theta_{min}, \theta, \delta, M_I, M_R) \geq \Delta\theta$}}
\STATE {$Relocate(v, \theta_{min}, \theta, \delta, M_I, M_R)$;}
\STATE add neighbors of $v$ to $Q$
\ENDIF
\ENDWHILE
\end{algorithmic}
\end{algorithm}

%\begin{algorithm}
%\caption{$FinalVertexRelocation(M_I, M_R, \delta, \theta)$}
%\label{alg:improving_quality}
%\begin{algorithmic}[1]
%\FOR{$i \leftarrow$ 1 to n}
%\FOR{each vertex $v \in M_R$}
%\STATE $\theta_{min}\leftarrow \angle_{min}(\triangle), \triangle \in \mathcal{N}(v)$;\COMMENT{$\angle_{min}(\triangle)$ is the min.~angle of $\triangle$, and $ \mathcal{N}(v)$ are the facets incident to $v$}
%\IF{{$RelocateImproves(v, \theta_{min}, \theta, \delta, M_I)$} }
%\STATE {$Relocate(v, \theta_{min}, \theta, \delta, M_I, M_R)$;}
%\ENDIF
%\ENDFOR
%\ENDFOR
%\end{algorithmic}
%\end{algorithm}

Optimal vertex positions would be found if no vertex could be relocated to a better position anymore. However, we empirically restrict the angle improvement $\Delta\theta=0.1^{\circ}$(cf.~Sec. \ref{sec:exp_post_subalgorithm}), since afterward there are usually no significant improvements anymore.

\section{Error Metric} 
\label{sec:local_error}

\subsection{Hausdorff Distance}
\label{sec:Hausdorff_definition}

We use the Hausdorff distance to measure the approximation error between $M_R$ and $M_I$. Let $d(p, q)$ denote the Euclidean distance between two points $p$ and $q$ in 3D space. The distance of a point $p$ to a surface $M$ is defined as the shortest distance between p and any point of $M$
\begin{equation}
\label{equ:1}
d(p, M) = \min_{q \in M}d(p, q).
\end{equation}
The one-sided Hausdorff distance from a source surface $M$ to a target surface $T$ is defined as the maximum of all such point to surface distances :
\begin{equation}
\label{equ:2}
d_h(M, T) = \max_{p \in M} d(p, T).
\end{equation}
The one-sided Hausdorff distance is in general not symmetric, i.e.~ $d_h(M,T) \neq d_h(T,M)$. 
It is easily possible to construct counter-intuitive cases where $d_h(M,T)=0$ but $d_h(T,M)$ is arbitrarily large\footnote{Choose $T$ as a sphere of radius $r$ and $M$ as a hemisphere subset. Then with $r\rightarrow \infty$ also $d_h(T,M)\rightarrow \infty$.}.\\

\noindent The two-sided Hausdorff distance~\cite{henrikson99} between $M$ and $T$ resolves this issue by symmetrization
\begin{equation}
\label{equ:3}
d_H(M, T) = \max\{d_h(M, T), d_h(T, M)\}.
\end{equation}

\subsection{Approximating $d_H$ with Stratified Sampling}
\label{sec:stratify_sampling}

The exact evaluation of the two-sided Hausdorff distance is computationally very expensive~\cite{Tang09}. However, by careful surface sampling in combination with local updates of shortest point-to-surface links it is possible to obtain an efficient yet sufficiently accurate approximation, as discussed next.\\

\noindent Assume that $M$ is sampled by a point set $S_M\subset M$.
Then the one-sided Hausdorff distance can be approximated by
\begin{equation}
\label{equ:4}
d_h(M, T) \approx \max_{a \in S_M}d(a, T).
\end{equation}
By additionally sampling $T$ we obtain an approximation of the two-sided Hausdorff distance
\begin{equation}
\label{equ:5}
d_H(M, T) \approx \max\left(\max_{a \in S_M}d(a, T),\max_{b \in S_T}d(b, M)\right),
\end{equation}
with $S_T$ being a set of point samples on $T$. Note that our approximation still measures the exact distance from sample points to the complete surface, which provides significantly higher accuracy than a point cloud distance $d_H(S_M,S_T)$. Moreover, it ensures that we strictly underestimate the real distance.
Following the triangle inequality, the approximation error of our sampled Hausdorff distance is bounded by $\max\{d_h(M,S_M), d_h(T,S_T)\}$, i.e. the maximum gap between sample points. Consequently, in order to guarantee a good approximation, we target a uniform sampling of the surfaces. 
However, for piecewise linear triangle meshes the maximal distance often occurs at creases or corners, i.e. at mesh edges or vertices. This two observations motivate our \textit{stratified sampling} approach, which is uniform on faces but additionally adds samples on edges and vertices, as a kind of greedy importance sampling.\\

\noindent\textbf{Sampling the facets}. 
Instead of uniformly sampling the complete surface, we empirically found that better results can be obtained by uniformly sampling per triangle. In this way we obtain an automatic adaptivity, since more samples are available in structurally complex and highly curved areas, where also more triangles are necessary to describe such shape. Since our meshes are often highly non-uniform, we add a slight local smoothing of the sampling density by

\begin{equation}
\label{equ:6}
n(f_i) = n_{f} \cdot  \frac{1+|\mathcal{N}_{f_i}|}{1 + \sum_{f_j \in \mathcal{N}_{f_i}}{\frac{\mathcal{A}_j}{\mathcal{A}_i}}},
\end{equation}
where $n_f$ is the average number of samples per facet specified by the user, $\mathcal{N}_{f_i}$ are the neighbor facets that share vertices with $f_i$, and $\mathcal{A}_i$ is the area of facet $f_i$. 
Choosing a large $n_f$ offers a tighter approximation of the Hausdorff distance, however, resulting in high computational complexity. In our experiments, we found that $n_f = 10$ gives a good tradeoff between efficiency and effectiveness (cf.~Fig.~\ref{fig:exp_local_error}). 

\begin{figure}[hbt]
    \centering
    \begin{minipage}{29mm}
        \centering
        \includegraphics[width=28mm]{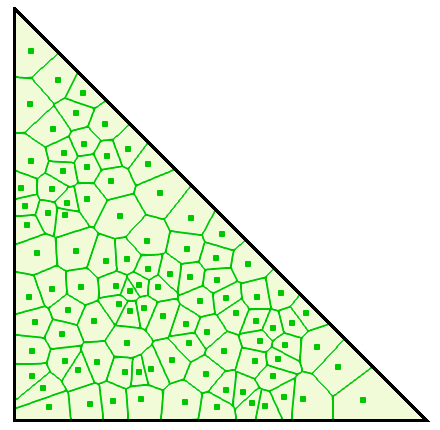}\\
    {\small (a)}
    \end{minipage}
    \begin{minipage}{29mm}
        \centering
        \includegraphics[width=28mm]{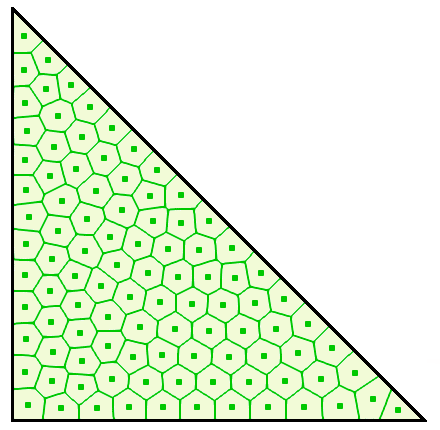}\\
    {\small (b)}
    \end{minipage}
    \begin{minipage}{29mm}
        \centering
        \includegraphics[width=28mm]{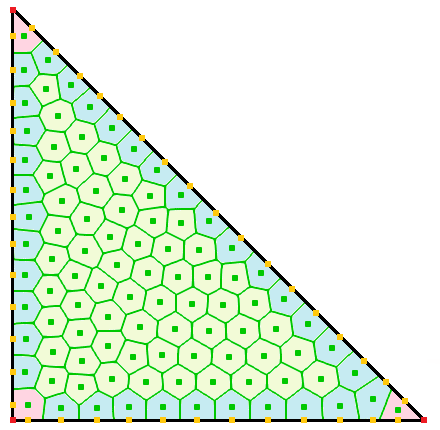}\\
    {\small (c)}
    \end{minipage}\\
    \caption{Stratified sampling process. (a) Initial facet samples; (b) Optimized facet samples with BVD; (c) Edge samples. The facet, edge, and vertex samples are rendered in green, yellow and red, respectively.}
\label{fig:triangle_samples}
\end{figure}

To distribute the samples evenly on a triangle $f_i$, we first generate $n(f_i)$ samples randomly, and then perform a Lloyd relaxation process onto a bounded Voronoi Diagram (BVD)~\cite{Tournois10} (Fig.~\ref{fig:triangle_samples}(b)). Usually, five iterations are sufficient to generate quasi-uniformly distributed samples on facets. 

\noindent\textbf{Sampling the edges}.
By counting the number of incident Voronoi cells to an edge we first estimate the local sampling density. The resulting number of samples is then evenly distributed along the edge (Fig~\ref{fig:triangle_samples}(c)). 

\noindent\textbf{Sampling the vertices}. The position of a vertex is simply regarded as its own vertex sample.
\begin{figure}[hbt]
    \centering
    \begin{minipage}{44mm}
        \centering
        \includegraphics[width=44mm]{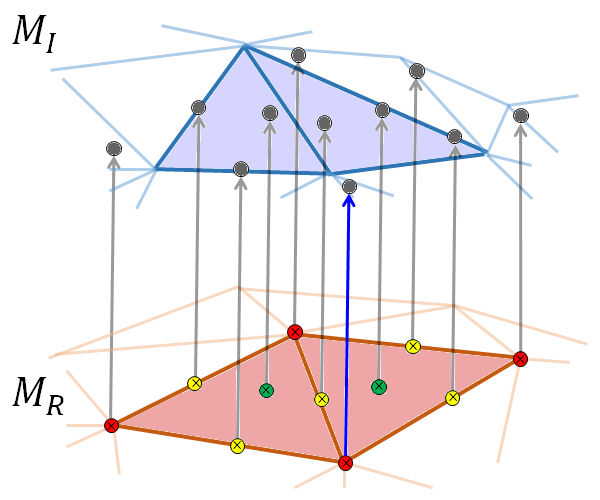}\\
    {\small (a) Stratified samples on $M_R$ and their shortest links to $M_I$.}
    \end{minipage}
    %\hspace{0.1in}
    \begin{minipage}{44mm}
        \centering
        \includegraphics[width=44mm]{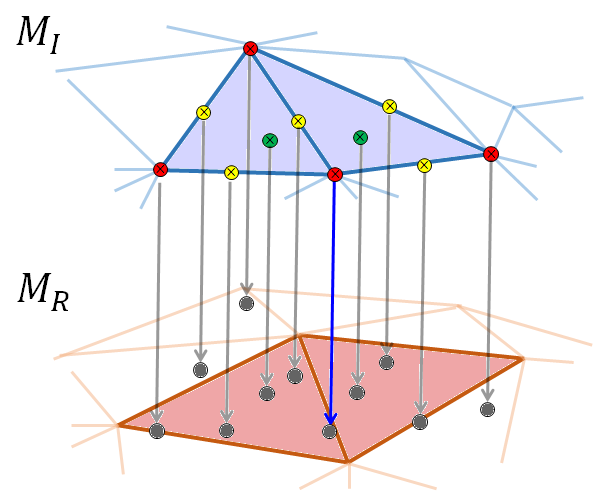}\\
    {\small (b) Stratified samples on $M_I$ and their shortest links to $M_R$.}
    \end{minipage}\\
    %\vskip0.1in
    \caption{Hausdorff distance approximation with stratified samples. Samples on vertices, edges, and faces are rendered in red, yellow, and green respectively. We use the circles centered with crosses to indicate the samples on one mesh, and the circles centered with dots to indicate their closest points on the other mesh. The approximated one-side Hausdorff distances are shown as the blue links, and the overall approximated Hausdorff distance is simply the length of the longer one.}
\label{fig:stratify_sample}
\end{figure}\\

By setting $S_I$ and $S_R$ as the stratified samples on $M_I$ and $M_R$, the Hausdorff distance between $M_I$ and $M_R$ is approximated using Eq.~(\ref{equ:5}), as illustrated by Fig.~\ref{fig:stratify_sample}.

\subsection{Local Update Scheme}
\label{sec:measure_scheme}
Each operator of our remeshing algorithm only changes a local area $L \subset M_R$ of the target mesh $M_R$ (cf.~Fig.~\ref{fig:operators} green area). Hence, it is possible to rapidly compute the Hausdorff distance for the updated mesh. In general both directions of the Hausdorff distance change and require an update.\\

\noindent\textbf{Updating $\mathbf{d_h(M_R,M_I)}$}: 
First of all notice that the one-sided Hausdorff distance can be decomposed w.r.t. to a modified local area $L\subset M_R$ into
$$d_h(M_R,M_I) = \max\{d_h(L, M_I), d_h(\bar{L}, M_I) \}$$
with $M_R=L \cup \bar{L}$ being a partition of $M_R$. Since $\bar{L}$ is unchanged, checking $d_h(L,M_I) \leq \delta$ is sufficient to verify that the modification does not violate the approximation error bound $\delta$. 
Thus, we first re-sample the modified local area $L$ with stratified samples $S_L$, and then efficiently evaluate the approximated Hausdorff distance $d_h(S_L,M_I)$ with a pre-computed axis aligned bounding box tree \cite{cgal-aabb} for the static mesh $M_I$.\\

\noindent\textbf{Updating $\mathbf{d_h(M_I,M_R)}$}: 
Checking the opposite direction is more difficult for two reasons. Firstly, since $M_R$ changes, we cannot simply pre-compute a static search tree. Secondly, decomposing the Hausdorff distance in the second argument is more intricate. In addition to $M_R$, $M_I$ must also be decomposed correctly. More specifically, we have $$d_h(M_I,M_R)\approx d_h(S_I,M_R) = \max\{ d_h(S_I^{L}, L), d_h(S_I^{\bar{L}}, \bar{L})\} $$
where $S_I = S_I^L \cup S_I^{\bar{L}}$ is a partitioning of samples on $M_I$ into those closer to $L$ and $\bar{L}$ (the rest of $M_R$) respectively.\\

Identifying the correct partitioning of samples $S_I$ would in general require global tests and is thus time consuming. Therefore, in order to enable a rapid update we only approximate this decomposition by tracking the history of samples in $S_I$. 
The key idea is to store for each sample $s_j \in S_I$ a link to its closest triangle $t_j \in M_R$ (Fig.~\ref{fig:stratify_sample}). After a local modification of $M_R$ these links will typically change in the vicinity of the modified area $L$. 
Thus, in order to avoid global checks, we only update the links of samples connecting to a region $L^+$, which enlarges $L$ by an additional ring of neighboring triangles (cf.~Fig.~\ref{fig:operators} gray area).
These new links can be efficiently found by constructing and querying an the axis aligned bounding box tree of the local area $L^+$. 
Note that the resulting approximated Hausdorff distance is a strict overestimator because shorter links to $\bar{L^+}$ might exist that are not investigated by our approximation. A positive side effect of our localized update scheme is that links cannot jump between geodesically far away regions, which could cause topologically inconsistent links.

\section{Implicit Feature Preservation}
\label{sec:feature_function_based_vertex_position}
\begin{figure}[hbt]
    \centering
    \begin{minipage}{20mm}
        \centering
        \includegraphics[width=19mm]{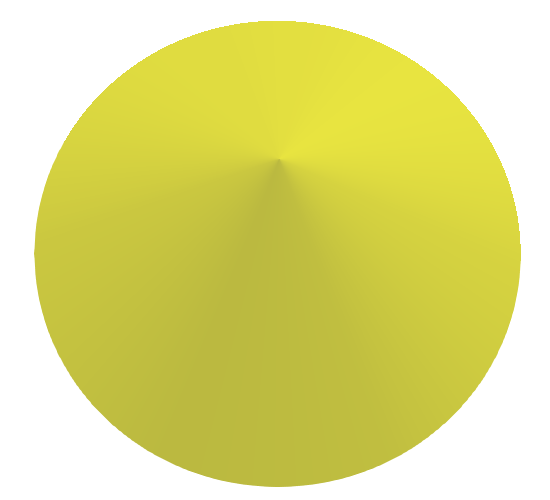}\\
    {\small (a) tip.}
    \end{minipage}
    \begin{minipage}{20mm}
        \centering
        \includegraphics[width=19mm]{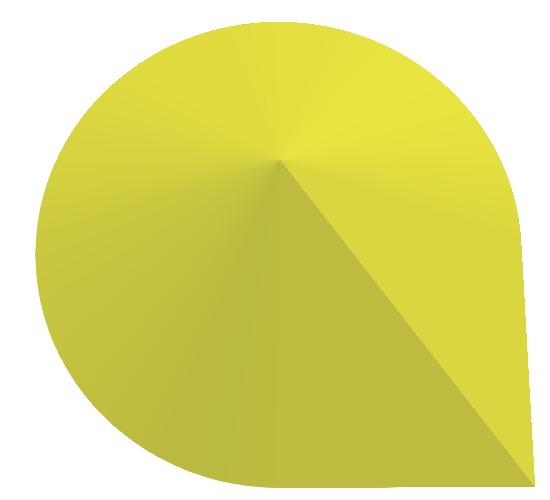}\\
    {\small (b) dart.}
    \end{minipage}
    \begin{minipage}{20mm}
        \centering
        \includegraphics[width=19mm]{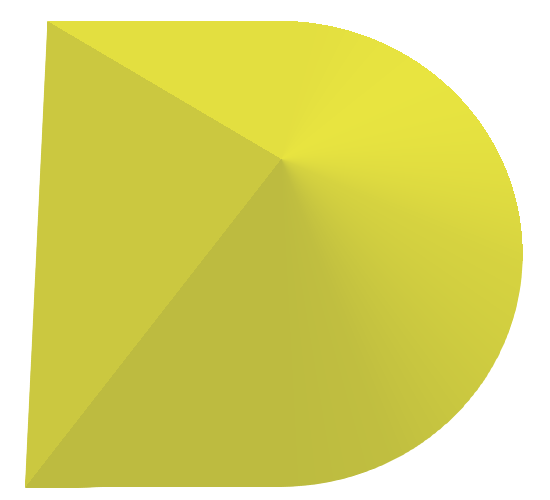}\\
    {\small (c) cusp.}
    \end{minipage}
    \begin{minipage}{20mm}
        \centering
        \includegraphics[width=19mm]{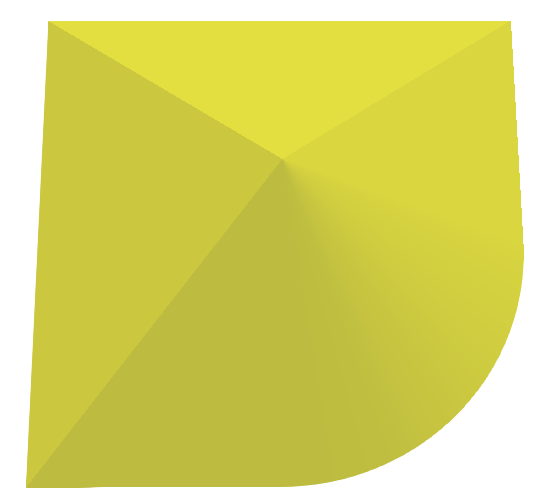}\\
    {\small (d) corner.}
    \end{minipage}\\
    \caption{Vertex features: (a) Tips, (b) Darts, (c) Cusps and (d) Corners are feature vertices who are adjacent to zero, one, two and three sharp creases respectively. } % \TODO{Kaimo}
\label{fig:features}
\end{figure}

A clean representation of geometric features as for instance sharp creases is important in many applications ranging from visualization to simulation. In a polygonal mesh, we distinguish between vertex and edge features. Vertex features include tips, darts, cusps and corners, while edge features are either creases or boundaries. Typical examples are depicted in Fig.~\ref{fig:features}. 

Instead of requiring an explicit tagging of features as most remeshing approaches, e.g.~\cite{Yan09, Yan13, Yan14A, Yan14B, Zhong13}, our goal is to implicitly preserve features. This not only releases users from the time-consuming burden of manual tagging but moreover often enables the recovery of features that were lost, e.g.~through inappropriate sampling by a 3D laser scanner.\\

In principle feature preservation could be simply a byproduct of approximation error minimization since incorrectly meshing features usually induces a large approximation error. Nevertheless, there are several reasons why special care is still required. First of all, the minimization of the approximation error is a non-convex problem such that bad initializations might lead to low-quality local minima, not well representing features. Moreover, anticipating feature locations and placing vertices accordingly will speed up the overall method. In our method special feature handling is done whenever vertices are either newly placed or relocated, i.e.~during (i) edge collapse, (ii) edge split and (iii) vertex relocation. Since robust and automatic feature detection is a difficult yet unsolved problem, we rely on a softer identification of features by means of a feature intensity function defined at vertices of the mesh.

\subsection{Feature Intensity}
\label{sec:feature_functions}
Feature vertices can be characterized by large \textit{Gaussian curvature} $\mathcal{K}(v)$, which in the discrete setting is identical to the angle defect, i.e.
\begin{equation}
\label{equ:7}
\mathcal{K}(v) = \left\{ \begin{array}{ll}
\pi - \theta_{sum}(v) & \textrm{v is on a boundary,}\\
2\pi - \theta_{sum}(v) & \textrm{otherwise},
\end{array} \right.
\end{equation}
where $\theta_{sum}(v)$ is the sum of interior angles adjacent to $v$.\\

Feature edges are characterized by large dihedral angles. Accordingly, for a vertex we define the \textit{feature edge intensity} $\mathcal{E}(v)$ to be the maximal unsigned dihedral angle of an edge adjacent to $v$
\begin{equation}
\label{equ:8}
\mathcal{E}(v) =  \max_{e \in \mathcal{N}_e(v)} |\mathcal{D}(e)|
\end{equation}
where $\mathcal{N}_e(v)$ are the edges adjacent to $v$ and $\mathcal{D}(e)$ is the dihedral angle at $e$.\\

\begin{figure}[hbt]
    \centering
    \begin{minipage}{29mm}
        \centering
        \includegraphics[width=29mm]{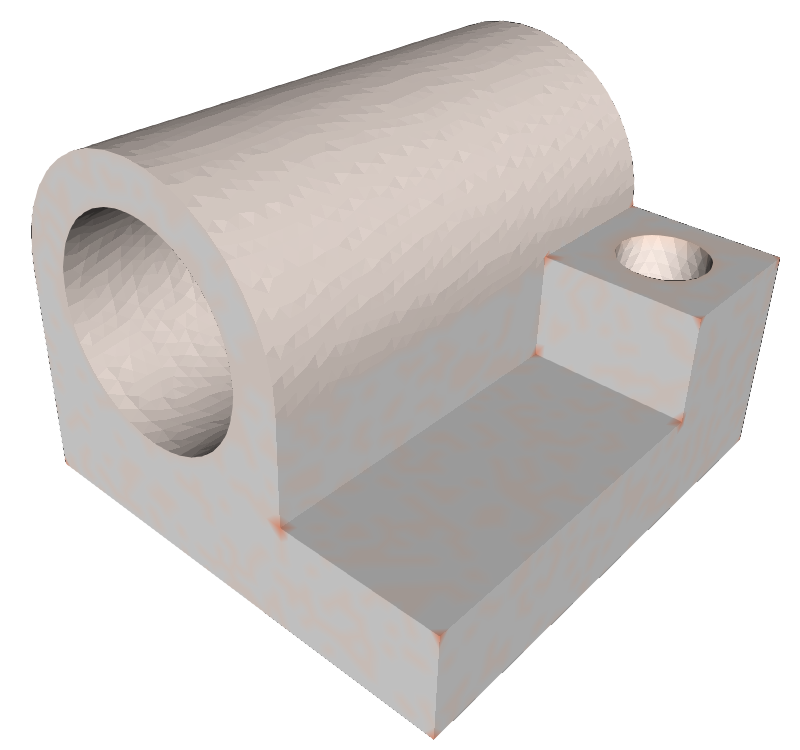}\\
    {\small $|{\mathcal{K}}(v)|$}
    \end{minipage}
    \begin{minipage}{29mm}
        \centering
        \includegraphics[width=29mm]{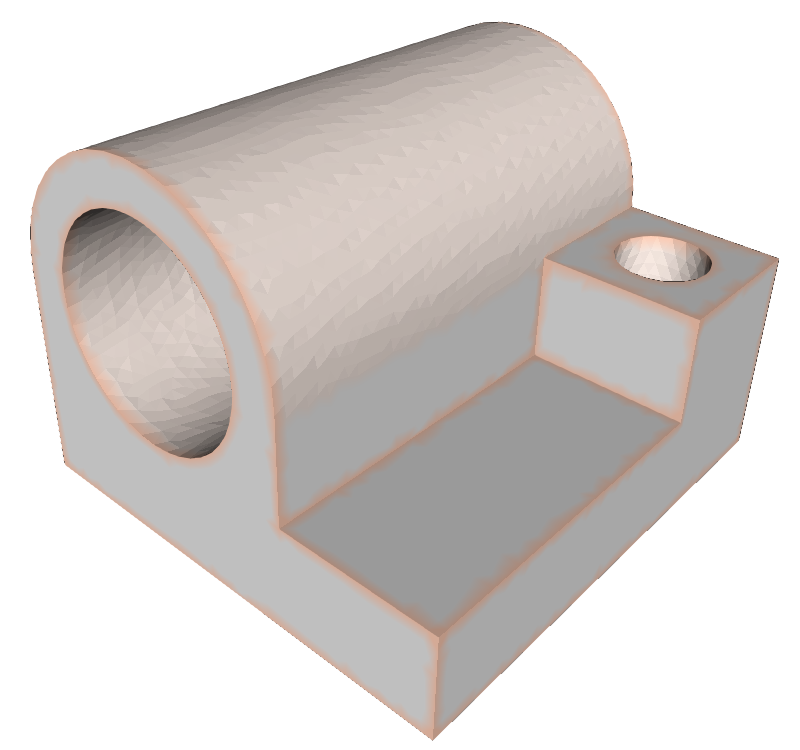}\\
    {\small ${\mathcal{E}}(v)$}
    \end{minipage}
    \begin{minipage}{29mm}
        \centering
        \includegraphics[width=29mm]{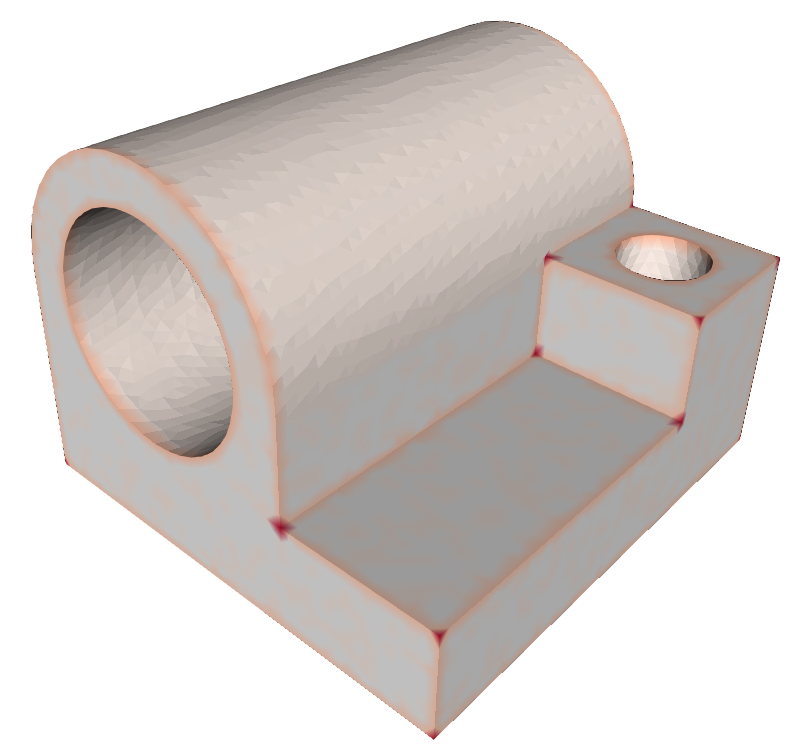}\\
    {\small ${\mathcal{F}}(v)$}
    \end{minipage}\\
    \caption{From left to right Gaussian curvature $|\mathcal{K}|$, feature edge intensity $\mathcal{E}$ and combined feature intensity $\mathcal{F}$, with higher intensities in red.}
\label{fig:feature_functions}
\end{figure}

Finally, the \textit{feature intensity} $\mathcal{F}(v)$ is defined as the combination
$$\mathcal{F}(v) = (\tau(|\mathcal{K}(v)|)+1)\cdot(\tau(\mathcal{E}(v))+1)-1$$
with the transfer function $\tau(x) = \min\{\pi, 2\cdot x\}$, which rescales values by a factor of $2$ and clamps them at $\pi$.
Thus, the feature intensity is a value between $0$ and $((\pi+1)^2-1)$ and corresponds to a \textit{logical or} whenever one of the individual intensities vanishes.
Fig.~\ref{fig:feature_functions} shows an examples of the three fields $|\mathcal{K}|$, $\mathcal{E}$ and $\mathcal{F}$.

\subsection{Feature-Sensitive Vertex Relocation}
\label{sec:feature_edge_collapse}
In our remeshing algorithm each local operation is combined with a subsequent relocation of the modified vertex in order to minimize the approximation error. This relocation of a single vertex is done in two stages. First a feature-sensitive initialization, specifically adapted to the local operation, and then a nonlinear minimization of the two-sided Hausdorff distance. A careful initialization of vertex positions is important to avoid poor local minima of the non-convex Hausdorff energy and additionally increases performance.\\

\noindent{\textbf{Edge Collapse Initialization}}: 
During an edge collapse of edge $e_{ij}$ the two vertices $v_i$ and $v_j$ merge into a new vertex $v_m$ and a new position for $v_m$ must be specified. 
One feature-sensitive standard technique uses error quadrics~\cite{Garland97, Valette08}. However, since we anyway perform a more accurate approximation error-driven relocation subsequently, a simpler and faster position initialization based on our feature intensity is sufficient. We distinguish two cases. If the edge is incident to a single strong feature we want that $v_m$ snaps onto this feature, meaning that the default behavior is to snap to the vertex $v_k$ with higher feature intensity, i.e.~$v_k=\argmax_{v\in\{ v_i, v_j\}} \mathcal{F}(v)$. An unclear situation arises, when both feature intensities are similar, i.e.~$\mathcal{F}(v_i)\approx \mathcal{F}(v_j)$. This happens either in regions without features or for edges along a crease, where it is reasonable to initialize $v_m$ to the edge midpoint.
We decide for the midpoint initialization based on a parameter $\omega$, whenever
%$$\frac{|\mathcal{F}(v_i)-\mathcal{F}(v_j)|}{\max(\mathcal{F}(v_i),\mathcal{F}(v_j))} < \omega$$
$$|\mathcal{F}(v_i)-\mathcal{F}(v_j)| < \omega \cdot \max(\mathcal{F}(v_i),\mathcal{F}(v_j))$$
with $\omega=0.15$ in all our examples.\\

\noindent{\textbf{Edge Split Initialization}}: Since an edge split does not change the geometry we simply always initialize the new point as the edge midpoint. Further improvement is achieved through a nonlinear minimization of the two-sided Hausdorff distance, as discussed next.\\

\begin{figure}[hbt]
    \centering
    \begin{minipage}{29mm}
        \centering
        \includegraphics[width=28mm]{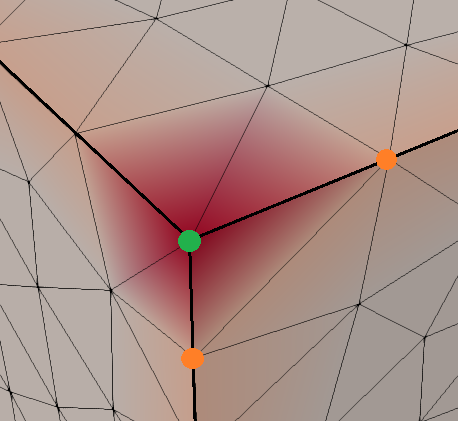}\\
    {\small (a)}
    \end{minipage}
    \begin{minipage}{29mm}
        \centering
        \includegraphics[width=28mm]{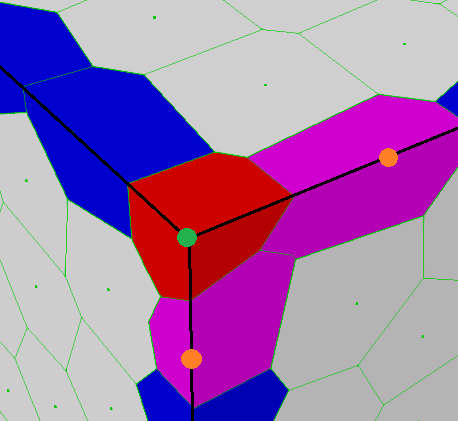}\\
    {\small (b)}
    \end{minipage}
    \begin{minipage}{29mm}
        \centering
        \includegraphics[width=28mm]{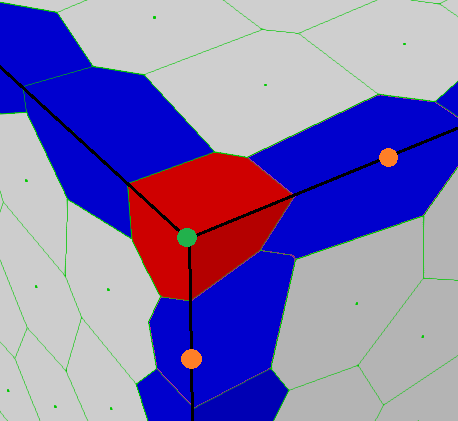}\\
    {\small (c)}
    \end{minipage}\\
    \caption{Visualization of $\mathcal{F}(v)$ in (a). The color of $v$'s Voronoi cell represents the number of important neighbors, with red=$0$, blue=$2$, purple=$3$ and gray=degree($v$) is shown in (b) and (c). 
    A wrong classification based on solely feature intensity in (b) is corrected by additionally checking importance of edges in (c).}
    %\caption{A special corner case in vertex relocates. The mesh are rendered with $\bar{\mathcal{F}}(v)$ values in (a), the Voronoi cells are shown in (b) and (c). The color of $v$'s Voronoi cell varies according to the number of its \textit{effective} neighboring vertices, where red, blue, purple and gray represent zero, two, three and $v$'s degree, respectively. When only the first item in Eq.~(\ref{equ:11}) is applied for evaluating $\mathcal{B}$, the crease vertices (orange) neighboring to a feature vertex (green) might be wrongly relocated (b). Appending the second item in Eq.~(\ref{equ:11}) perfectly solves this problem (c).}
\label{fig:corner}
\end{figure}

\noindent{\textbf{Vertex Relocation Initialization}}: Without topological mesh modifications (edge collapse or split), a vertex relocation is initialized in the following way. The goal is to anticipate a good relocation position for $v$ while preserving feature corners and creases. Therefore, we first classify the vertex $v$ as either being (i) a feature vertex, (ii) a crease vertex or (iii) a smooth vertex. If $v$ is a feature vertex it simply remains at its position. If it is on a crease, we relocate $v$ to the CVT barycenter of its two neighboring crease vertices. And only if it is a smooth vertex we relocate it to the CVT barycenter of all one-ring neighbors. The classification is done by counting how many neighbors $v_i$ are of similar or higher importance as $v$, i.e.~if $ \mathcal{F}(v_i) \geq \zeta \cdot \mathcal{F}(v)$ for a tolerance parameter $\zeta \in (0,1)$. In this way $v$ is classified as (feature vertex/crease vertex/smooth vertex), depending on whether (none/two/all) of its neighbors are of similar feature intensity.
Unclear cases where $k$ out of $n$ neighbors are of similar importance are classified towards the closer possibility, meaning as a crease vertex if $k$ is closer to $2$ or as a smooth vertex if $k$ is closer to $n$. 
Wrong classifications where two crease vertices are connected by a non-crease edge can be avoided by only counting important neighbors that are connected by an important edge with $|D(e_{v,v_i}|+1 \geq \zeta \cdot (\mathcal{E}(v) +1)$, as illustrated in Fig.~\ref{fig:corner}. 

\begin{figure}[hbt]
    \centering
    \begin{minipage}{29mm}
        \centering
        \includegraphics[width=29mm]{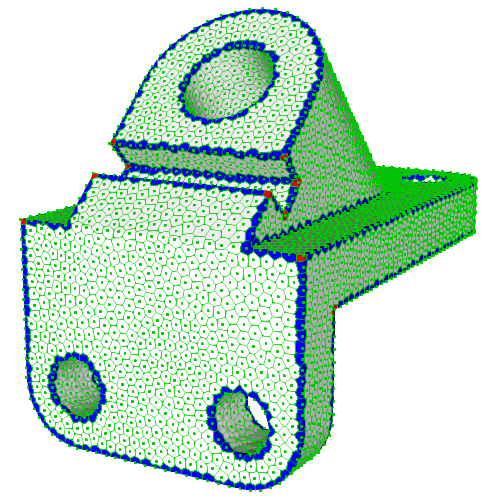}\\
    %{\small $\bar{\mathcal{F}}_s$}
    \end{minipage}
    \begin{minipage}{29mm}
        \centering
        \includegraphics[width=29mm]{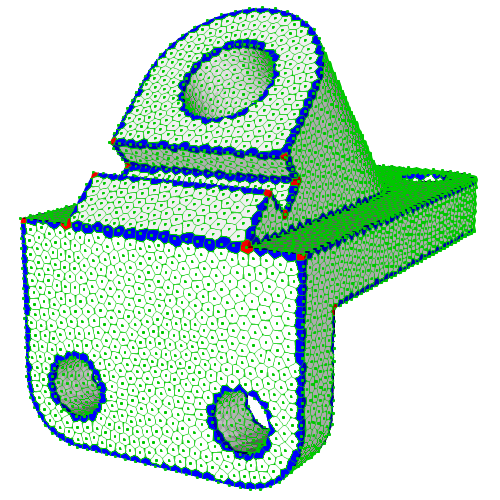}\\
    %{\small $\bar{\mathcal{F}}_m$}
    \end{minipage}
    \begin{minipage}{29mm}
        \centering
        \includegraphics[width=29mm]{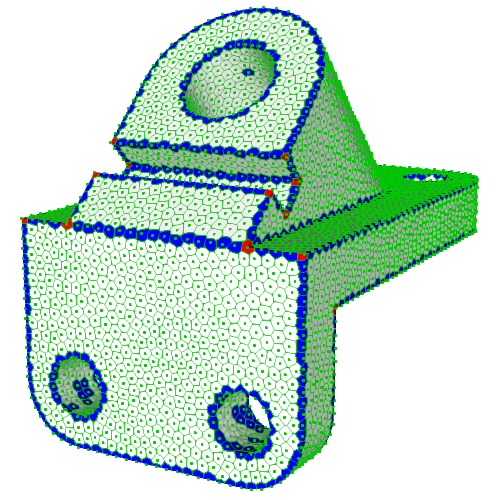}\\
    %{\small $min(\bar{\mathcal{F}}_s,\bar{\mathcal{F}}_m)$}
    \end{minipage}\\
    \vskip0.1in
    \begin{minipage}{29mm}
        \centering
        \includegraphics[width=29mm]{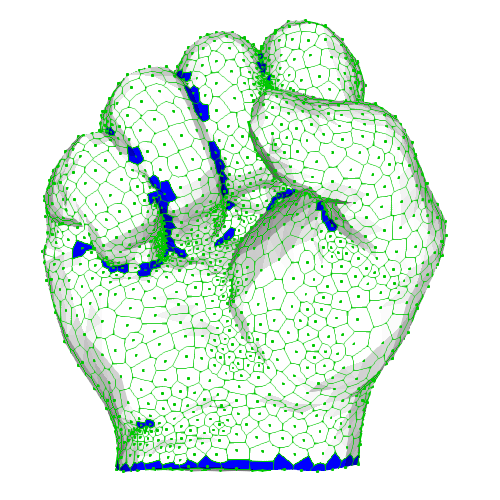}\\
    {\small $\zeta=0.3$.}
    \end{minipage}
    \begin{minipage}{29mm}
        \centering
        \includegraphics[width=29mm]{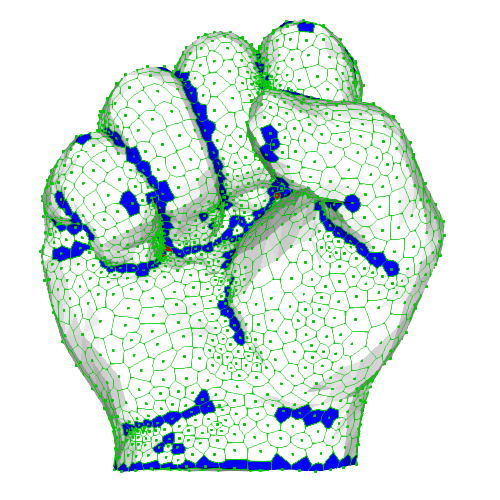}\\
    {\small $\zeta=0.5$.}
    \end{minipage}
    \begin{minipage}{29mm}
        \centering
        \includegraphics[width=29mm]{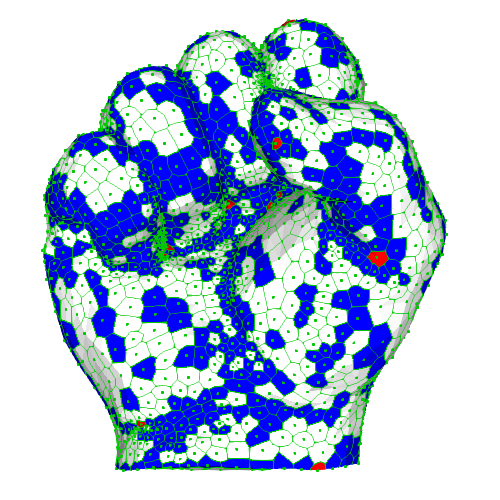}\\
    {\small $\zeta=0.7$.}
    \end{minipage}
    \caption{Influence of the classification tolerance $\zeta$. The Voronoi cells of vertices color code the classification with red=feature vertex, blue=crease vertex and white=smooth vertex. We show the Anchor model with obvious features (top) and the Hand model without obvious features (bottom).}
\label{fig:feature_relocate_type}
\end{figure}

The parameter $\zeta$ controls the feature classification. For higher values of $\zeta$ more vertices are implicitly treated as features and thus prevented from free movement. 
Fig.~\ref{fig:feature_relocate_type} shows an example classification for $\zeta$ varying between $0.3$ and $0.7$. In our experiments the default is $\zeta=0.5$
%When $\zeta = 0$, our relocate strategy degenerates to the original non feature preserving case~\cite{Botsch04}. According to different requirements for feature preservation, $\delta_r$ can be set between $30\%$ and $70\%$, as shown in Fig.~\ref{fig:feature_relocate_type}. In our experiments, we set $\delta_r$ as $50\%$ by default.
%Fig.~\ref{fig:feature_relocate_type} indicates that the saliency functions based metric does not only preserve sharp features robustly, such as the CAD models, but also provide a flexible way for preserving unconspicuous features, such as the hand model.
\\

%\subsection{Vertex Relocation}
%\label{sec:feature_vertex_relocate}

%Table \ref{tab:vertex_types} summarizes the classification and explains the classification of unclear cases.
%\begin{table}[hbt]
%    \caption{Classification of vertices.}
%    \label{tab:vertex_types}
%    \centering
%    \begin{threeparttable}
%    \begin{tabular}{|c|lll|}
%        \hline
%\textbf{Case} & $\bm{\#(\mathcal{B}(v_i, v))}$ & \textbf{Implication} & \textbf{Relocation strategy}\\
%        \hline
%1 & $0$  		& feature vertex 			& remain unchanged\\
%2 & $2$  		& crease vertex 				& to the CVT barycenter\\
%3 & $Degree(v)$	& smooth vertex 				& to the CVT barycenter\\
%4 & $others$	& unclear 			& round to case 2 or 3 *\\
%        \hline
%    \end{tabular}
%    \begin{tablenotes}
%	\item[*] Let $\#(\mathcal{B}(v_i, v))=n$. If $|n-2| \geq |n-Degree(v)|$, then $v$ is treated as a free vertex, the same as case 3; otherwise, we select two vertices $v_i, v_j \in \mathcal{N}(v)$ who possess the highest $\bar{\mathcal{F}}$ values, and relocate $v$ to the CVT barycenter of $v_i$ and $v_j$, similar to case 2.
%	\end{tablenotes}
%	\end{threeparttable}
%\end{table}

\noindent{\textbf{Nonlinear Hausdorff Distance Minimization}}: After one of the former initializations is done, we further optimize the position of a vertex $v$ by directly minimizing the approximate two-sided Hausdorff distance of Sec. \ref{sec:stratify_sampling}. This optimization is highly nonlinear, since changing the position of $v$ changes the samples $S_R$ of the modified mesh $M_R$ as well as the links from the input mesh samples $S_I$ to $M_R$. We perform an optimization similar to the Hausdorff distance minimization technique proposed by Winkler et al.~\cite{Winkler08}. However, we improve the feature-sensitivity by adjusting the weighting scheme with our feature intensity function $\mathcal{F}$, as detailed next.\\

We want to relocate $v$ in order to minimize the two-sided Hausdorff distance between the local area $L\subset M_R$ consisting of all one-ring triangles of $v$ and the subregion of $M_I$ with links into $L$, referred to as $M_{I \rightarrow L}$. Assume that the corresponding subsets of samples for our Hausdorff distance approximation are $S_L \subset S_R$ and $S_{I\rightarrow L} \subset S_I$.
Then according to Motzkin and Walsh's theorem~\cite{Motzkin59} there exist weights $w_i$ and $\hat{w}_i$ such that minimizing Eq.~(\ref{equ:5}) w.r.t.~the center vertex $v$ is equivalent to minimizing
\begin{equation}
\label{equ:12}
\sum_{a_i \in S_L} w_i | a_i(v)-\hat{a}_i|^2 + \sum_{b_i \in S_{I\rightarrow L}} \hat{w}_i | \hat{b}_i(v) - b_i |^2
\end{equation}
where $\hat{a}_i\in M_{I\rightarrow L}$ and $\hat{b}_i(v)\in L$ are the closest points to $a_i$ and $b_i$ respectively. Freezing the closest point pairs and assuming a linear barycentric relation 
$$a_i(v)-\hat{a}=\alpha_i v + \beta_i d_i + \gamma_i e_i -\hat{a}= \alpha_i v - p_i$$ 
with constant $p_i = \hat{a}-\beta_i d_i - \gamma_i e_i$ and similarly expressed
$\hat{b}_i(v) - b_i = \hat{\alpha}_i v - \hat{p}_i$, the optimal position $v^*$ can be computed analytically 
%by setting the derivative of Eqn.~(\ref{equ:12}) to zero
%$$ v^* = \frac{\sum_{a_i \in S_L} w_i \alpha_i p_i + \sum_{b_i \in S_{I\rightarrow L}} \hat{w}_i \hat{\alpha}_i \hat{p}_i}{\sum_{a_i \in S_L} w_i \alpha_i^2 + \sum_{b_i \in S_{I\rightarrow L}} \hat{w}_i \hat{\alpha}_i^2 }$$
\begin{equation} 
v^* = \frac{\sum_{a_i} w_i \alpha_i p_i + \sum_{b_i} \hat{w}_i \hat{\alpha}_i \hat{p}_i}{\sum_{a_i } w_i \alpha_i^2 + \sum_{b_i } \hat{w}_i \hat{\alpha}_i^2 }
\label{equ:13}
\end{equation}

\begin{figure}[t]
    \centering
    \begin{minipage}{29mm}
        \centering
        \includegraphics[width=28mm]{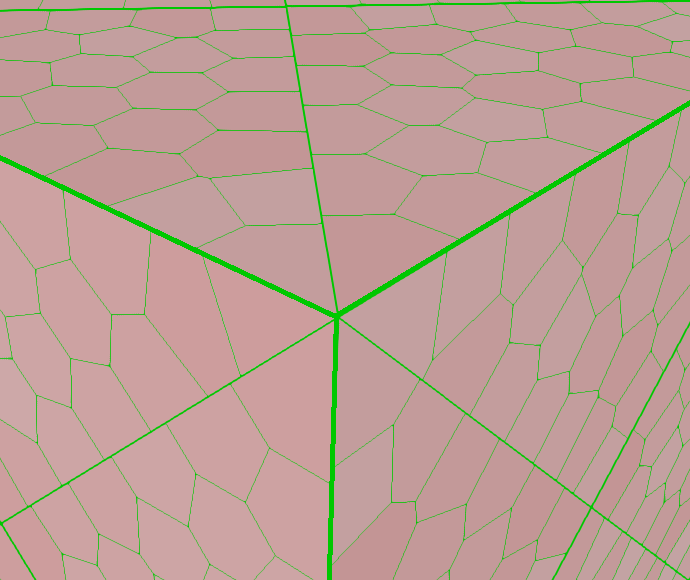}\\
    {\small $\mathcal{V}(a_i)$}
    \end{minipage}
    \begin{minipage}{29mm}
        \centering
        \includegraphics[width=28mm]{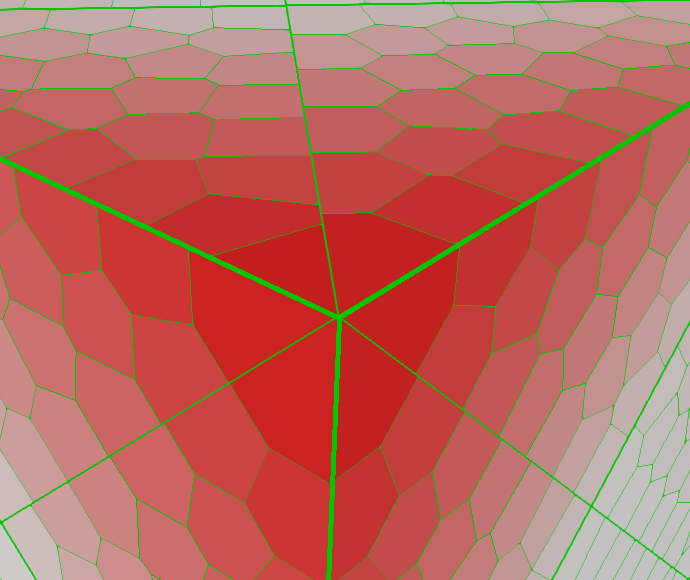}\\
    {\small $\mathcal{F}(a_i)$}
    \end{minipage}
    \begin{minipage}{29mm}
        \centering
        \includegraphics[width=28mm]{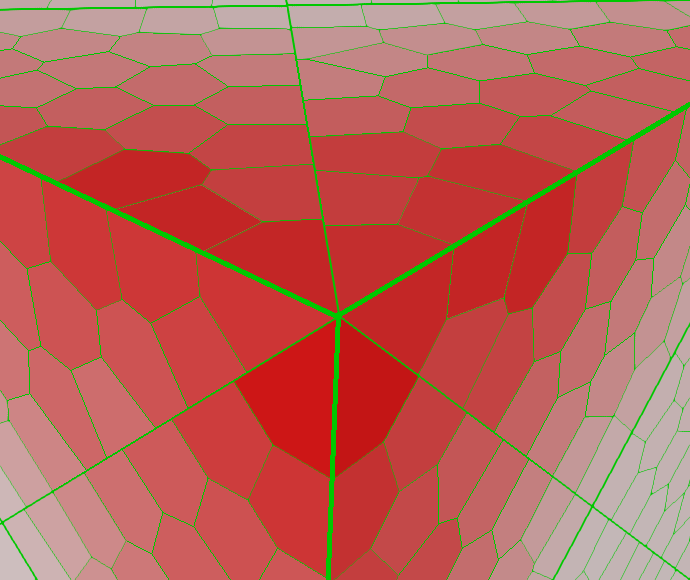}\\
    {\small $\mathcal{V}(a_i) \cdot \mathcal{F}(a_i)$}
    \end{minipage}\\
    \caption{Sample weights of a cube corner, with the integration of $\mathcal{V}(a_i)$ and $\mathcal{F}(a_i)$. Higher intensity of red highlights larger weight values.}
\label{fig:sample_weights}
\end{figure}

To find optimal weights $w_i$ and $\hat{w}_i$, Winkler et al.~\cite{Winkler08} use Lawson's algorithm~\cite{Lawson61} with iterative updates of the form
\begin{equation}
\label{equ:14}
w_i^{(k+1)}=w_i^{(k)} \cdot d(a_i^{(k)}, \hat{a}_i^{(k)}),
\end{equation}
where $d(a_i^{(k)}, \hat{a}_i^{(k)})$ is the Euclidean distance of the closest-point pair $(a_i, \hat{a}_i)$ after the $k$-th iteration, and initialization $w_i^{(0)}=1$. 
The idea behind this scheme is that samples with larger distances get a higher weight in the next iteration. Based on our feature intensity function and a sample density estimation, the weight update can be improved to
\begin{equation}
\label{equ:15}
w_i^{(k+1)}=w_i^{(k)} \cdot d(a_i^{(k)}, \hat{a}_i^{(k)}) \cdot \mathcal{V}(a_i^{(k)}) \cdot \mathcal{F}(a_i^{(k)})
\end{equation}
where $\mathcal{V}(a_i)$ is the Voronoi cell area of sample $a_i$, and $\mathcal{F}(a_i)$ is the linearly interpolated feature intensity value of $a_i$. Fig.~\ref{fig:sample_weights} illustrates the additional weighting factors. Advantages are a better feature preservation, improved robustness w.r.t.~non-uniform sampling and faster convergence.
%With the integration of $\mathcal{V}(a_i)$ and $\bar{\mathcal{F}}(a_i)$, the samples nearing to features or with low sampling densities are assigned higher weights, making the vertex position optimization unbiased to sampling densities and features are better preserved (Fig.~\ref{fig:sample_weights}).
\\

Each time when a local operator is applied, we iteratively optimize the new position of the effected vertex by minimizing Eq.~(\ref{equ:12}) using the new weighting defined in Eq.~(\ref{equ:15}). The optimization procedure is similar to the Expectation-maximization (EM) algorithm: in each iteration, we first calculate the optimal position $v^*$ of vertex $v$ using Eq.~(\ref{equ:13}), and then move $v$ to $v + \lambda(v^{*} - v), \lambda \in (0,1]$ with default $\lambda=0.9$ and update the closest point pairs. In practice, a few iterations usually suffice to get very close to the optimum.

\section{Experiment Results and Discussions}
\label{sec:experiment results}

We implemented our approach in C++ and tested on a 64-bit Windows 8.1 operating system. The CGAL library~\cite{Cgal} provided most of the basic data structures and operations. Timings for all the examples were conducted on a single 3.40GHz Intel(R) Core(TM) i7-2600K CPU with a 16GB RAM. We provide next a complete evaluation of our algorithm and comparisons with state-of-the-art approaches.

\subsection{Evaluation of the Local Error Update Scheme}
\label{sec:exp_local_error_measure_scheme}

\begin{figure}[hbt]
    \centering
    \begin{minipage}{44mm}
        \centering
        \includegraphics[width=44mm]{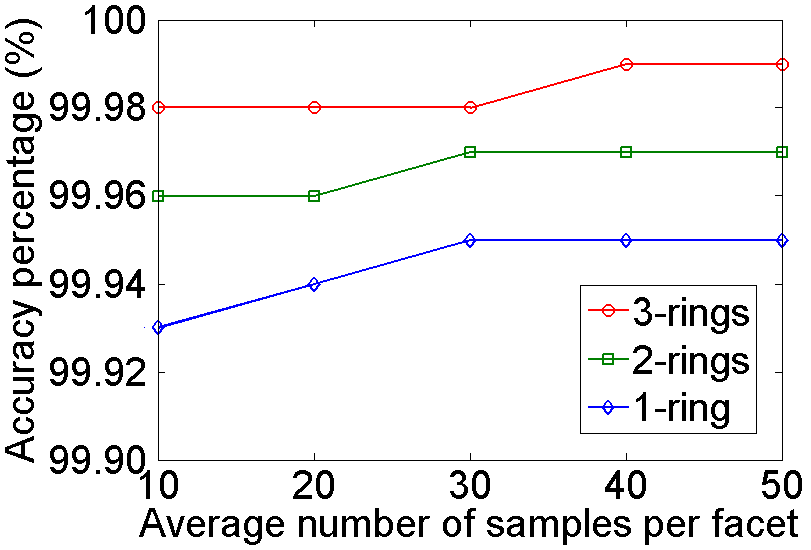}\\
    %{\small a. To be added.}
    \end{minipage}
    %\hspace{0.1in}
    \begin{minipage}{44mm}
        \centering
        \includegraphics[width=44mm]{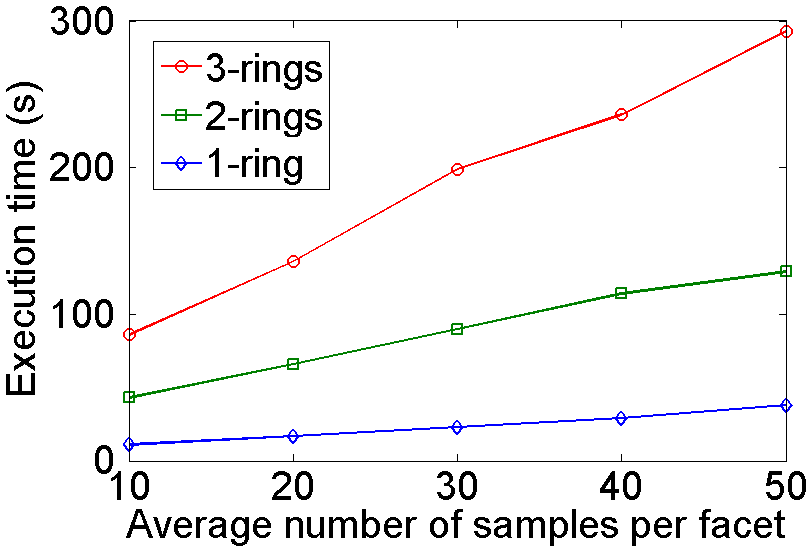}\\
    %{\small b. To be added.}
    \end{minipage}\\
    %\vskip0.1in
    \caption{The nearest point searching accuracy and execution time of our local error scheme, with respect to different sampling densities and inner patch sizes. In this experiment, $\delta=0.2\%$ of the bounding box's diagonal length($\%bb$) and $\theta=30^{\circ}$. The above data are the averages of 10 consecutive executions with the Rockerarm model (Table~\ref{table:comparison}) as input.}
\label{fig:exp_local_error}
\end{figure}

For efficient local error update we use the axis aligned bounding box tree of CGAL. As global error update is too compute-intensive we only verify how the inner patch sizes and sampling densities affect the effectiveness and efficiency of the local error update scheme (Fig.~\ref{fig:exp_local_error}). We find that even when the inner patch size is set to one-ring (cf. Fig.~\ref{fig:operators}), our local error update approach catches more than $99.9\%$ of the global nearest points. The accuracy increases little with higher sampling density and larger inner patch size. We set the inner patch sizes as one-ring facets and the average sampling number per facet as ten in all experiments.

\subsection{Evaluation of Vertex Position Optimization}
\label{sec:exp_vertex_position_optimization}
\begin{figure}[hbt]
    \centering
    \begin{minipage}{29mm}
        \centering
        \includegraphics[width=28mm]{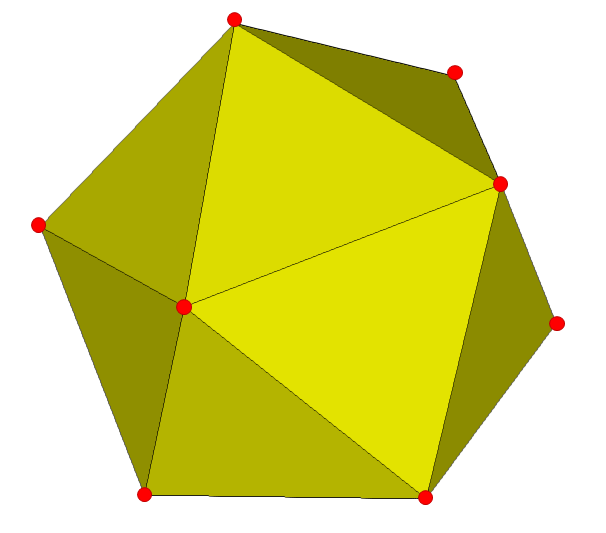}\\
    {\small (a) Icosahedron}
    \end{minipage}
    \begin{minipage}{29mm}
        \centering
        \includegraphics[width=28mm]{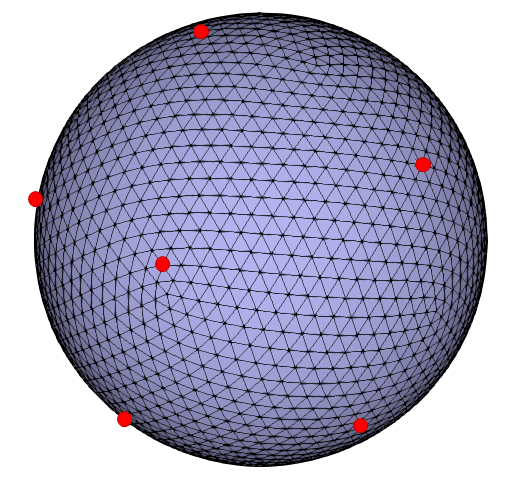}\\
    {\small (b) Interpolation}
    \end{minipage}
    \begin{minipage}{29mm}
        \centering
        \includegraphics[width=28mm]{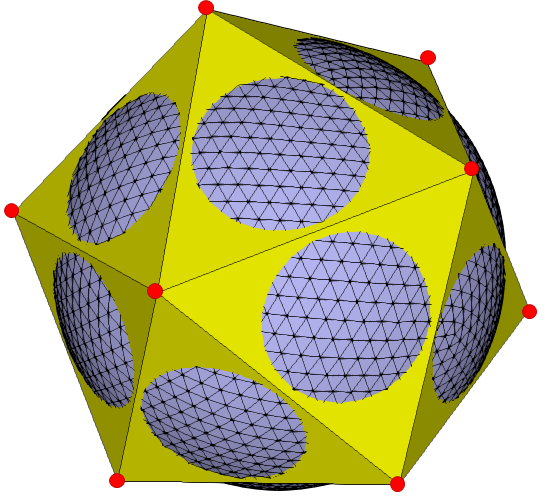}\\
    {\small (c) Approximation}
    \end{minipage}\\
    \caption{Demonstration of interpolation and approximation. We use an icosahedron (a) to interpolate and approximate a sphere ((b) and (c)). In interpolation, the vertices of $M_R$ are guaranteed to be on the surface of $M_R$, while in approximation, the error is minimized, regardless of whether the vertices of $M_R$ are on the surface of $M_I$.}
\label{fig:exp_interpolation_approximation}
\end{figure}

Solely applying the vertex position initialization makes most vertices of $M_R$ stay near or on the surface of $M_I$. Though this quasi-interpolation preserves sharp features, the optimal geometric fidelity is usually achieved when $M_R$ is an approximation of $M_I$. We visually demonstrate the difference between interpolation and approximation in Fig.~\ref{fig:exp_interpolation_approximation}. In the interpolation case, the Hausdorff distance (Hdist)~\cite{cignoni98} between the sphere and the icosahedron is $6.48(\%bb)$, and the root mean square (RMS) distance is $4.84(\%bb)$; while in the approximation case, the Hdist and RMS distances between them are $5.00(\%bb)$ and $1.09(\%bb)$, respectively.

%To further reduce the approximated errors between $M_I$ and $M_R$, we iteratively relocate the positions of vertices after applying each local operator; To better preserve the features in $M_R$, the saliency function measures are integrated, as described in Sec.~\ref{sec:position_optimization}.

\begin{figure}[hbt]
    \centering
    \begin{minipage}{44mm}
        \centering
        \includegraphics[width=44mm]{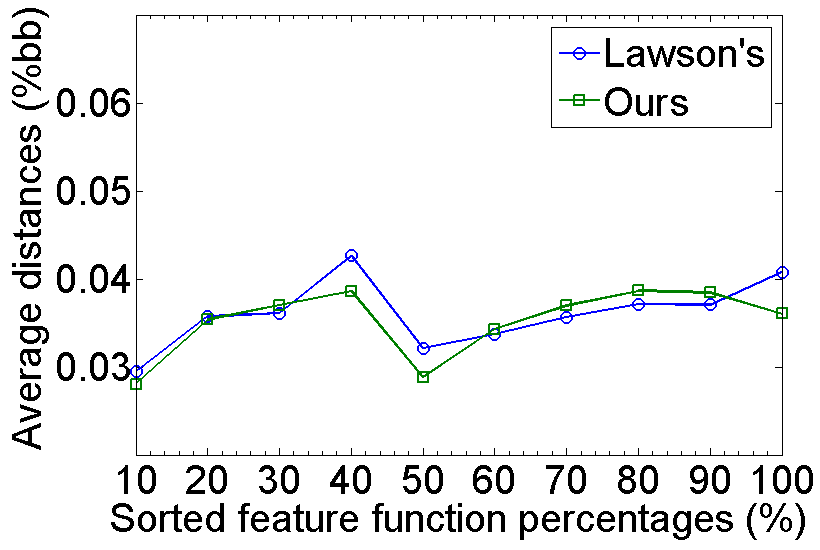}\\
    %{\small a. To be added.}
    \end{minipage}
    %\hspace{0.1in}
    \begin{minipage}{44mm}
        \centering
        \includegraphics[width=44mm]{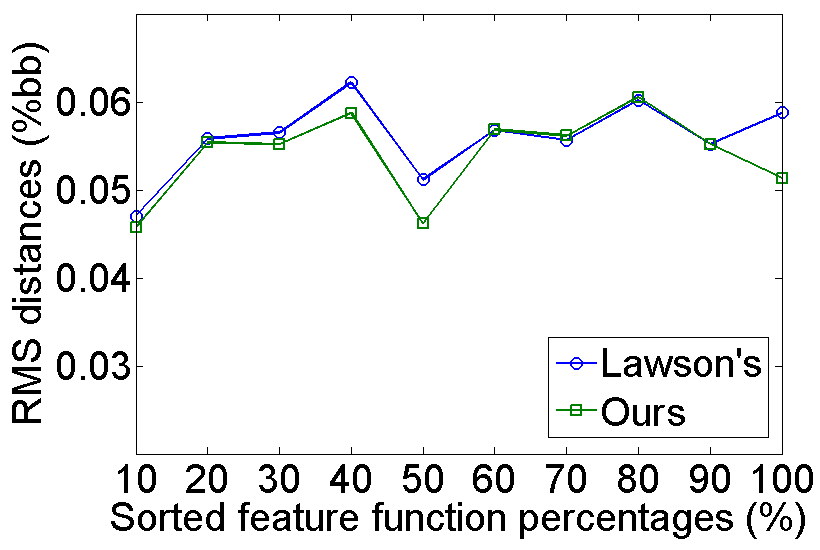}\\
    %{\small b. To be added.}
    \end{minipage}\\
    %\vskip0.1in
    \caption{Comparison of geometric fidelity between Lawson's weighing scheme~\cite{Winkler08} and ours(Eq.~\ref{equ:15}). We sort the vertices in $M_I$ according to their saliency function values in ascending order, and compute their average distance and RMS distance in each bin. In this experiment, $\delta=0.2(\%bb)$, and $\theta=30^{\circ}$. The above data are the averages of 10 consecutive executions with the Hand model (Fig.~\ref{fig:feature_relocate_type}) as input.}
\label{fig:exp_weight_comparison}
\end{figure}

However, the approximation might destroy features when minimizing the local sample pair distances~\cite{Winkler08}. 
%By integrating saliency functions and Voronoi cell areas in the weighting scheme, our method reduces the approximation error and preserves features at the same time. 
Fig.~\ref{fig:exp_weight_comparison} compares the average distance and RMS distance based on Lawson's weighting scheme (Eq.~(\ref{equ:14})) and our improved weighting scheme (Eq.~(\ref{equ:15})). Generally, our weighting scheme reduces the average distance and RMS distance about $2.3\%$ and $3.1\%$ respectively. However, it reduces the approximation error of vertices on sharp features about $11.8\%$ and $12.8\%$, respectively. Therefore, applying the improved weighting scheme does not only reduce the approximation error, but also better preserves sharp features.

\begin{figure}[hbt]
\centering
\includegraphics[width= 0.95\linewidth]{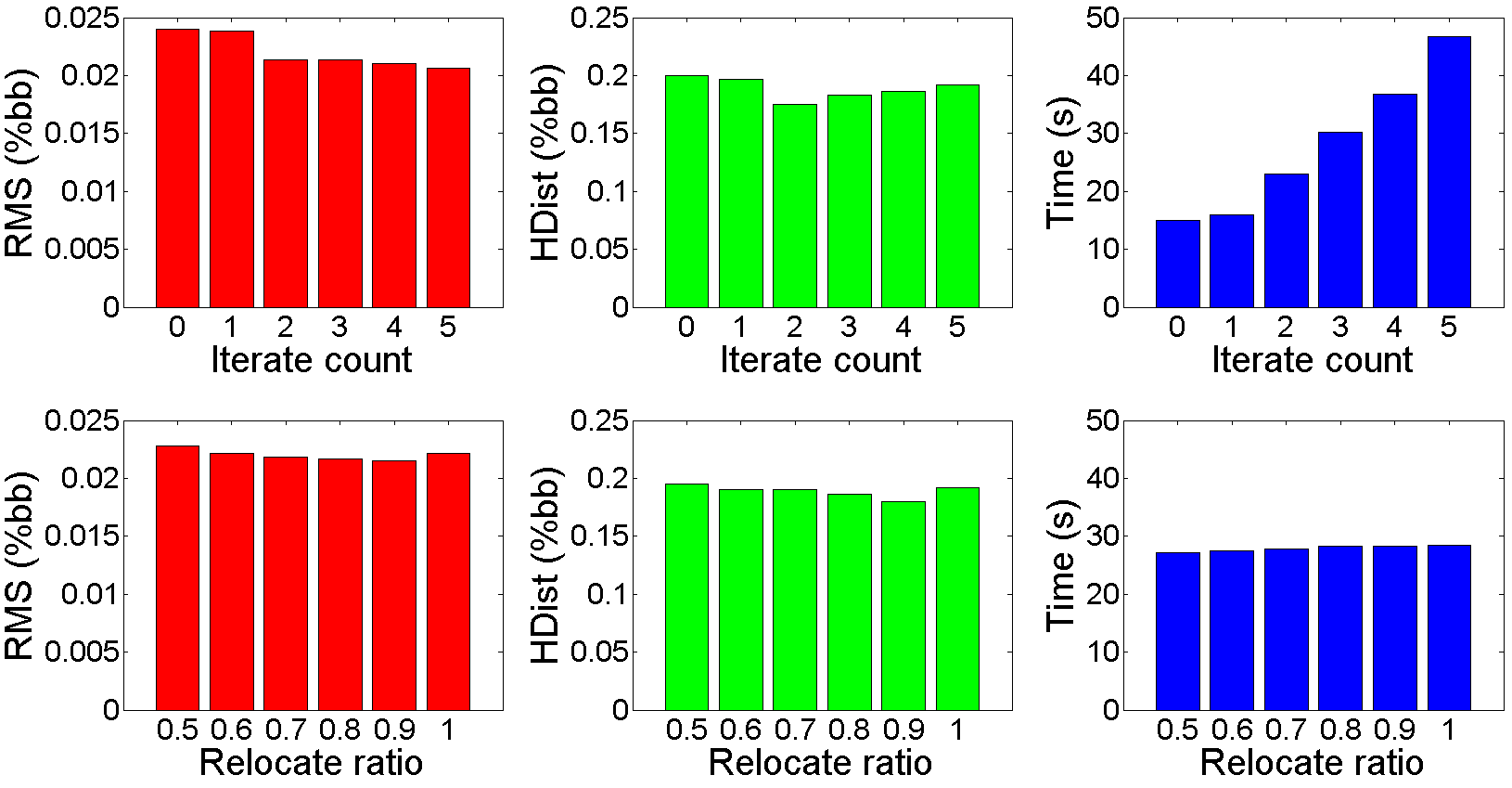}
\caption{Effectiveness of the iteration count and relocate ratio. For separability, RMS and Hdist bars are plotted in different scales. We set $\delta=0.2(\%bb)$, and $\theta=30^{\circ}$. The above data are the averages of 10 consecutive executions with the Homer model (Table.~\ref{table:comparison}) as input.}
\label{fig:exp_vertex_optimization_parameters}
\end{figure}

We extensively tested how the iteration count and relocate ratio ($\lambda$) affect the approximation error and execution time in the vertex position optimization procedure (in Fig.~\ref{fig:exp_vertex_optimization_parameters}).
%Note that when the iteration count is set to 0, the vertex position optimization is not applied, which yields the highest approximation error and the least execution time. The more iterations we apply, the more execution time is required.
We found that setting the iteration count to two and the relocate ratio to $0.9$ achieved the best compromise between effectiveness and efficiency. This configuration is used in all our later experiments.

\subsection{Evaluation of Initial Mesh Simplification}
\label{sec:exp_pre_subalgorithm}
\begin{figure}[hbt]
    \centering
    \begin{minipage}{29mm}
        \centering
        \includegraphics[width=28mm]{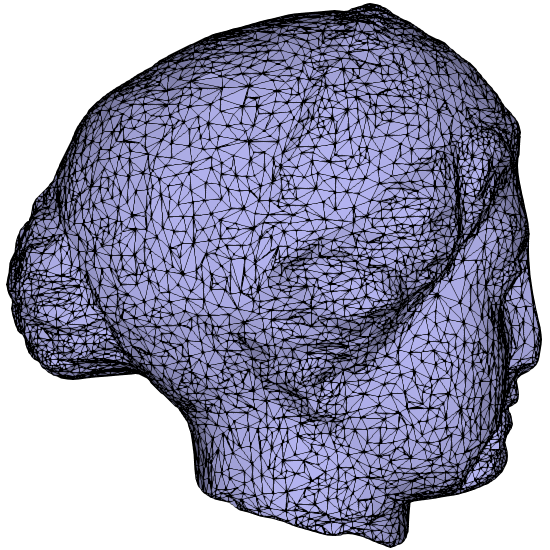}\\
    {\small (a)}
    \end{minipage}
    \begin{minipage}{29mm}
        \centering
        \includegraphics[width=28mm]{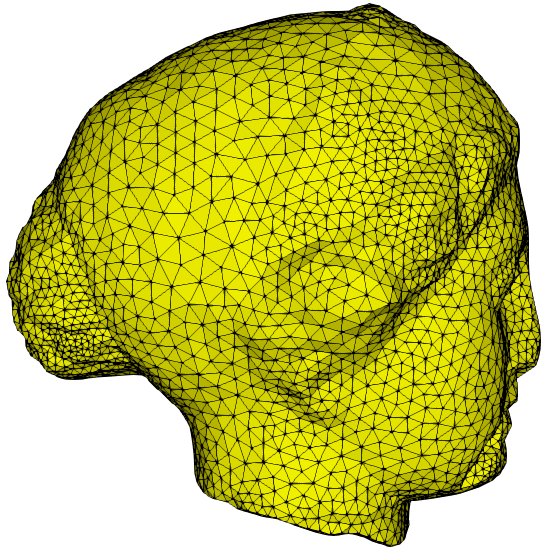}\\
    {\small (b)}
    \end{minipage}
    \begin{minipage}{29mm}
        \centering
        \includegraphics[width=28mm]{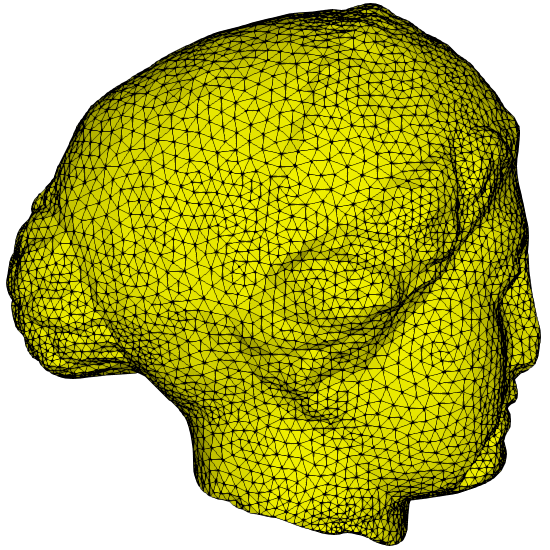}\\
    {\small (c)}
    \end{minipage}\\
    \caption{Egea models with/without executing Alg.~\ref{alg:reducing_complexity}. we set $\theta=40^{\circ}$ and $\delta=0.20(\%bb)$. The input (a) has 8.3k vertices. The result with Alg.~\ref{alg:reducing_complexity} enabled has 4.2k vertices (b), and spends 330 seconds; the result with Alg.~\ref{alg:reducing_complexity} disabled has 7.6k vertices (c), and spends 133 seconds.}
\label{fig:exp_mesh_simplification}
\end{figure}

In general, applying Alg.~\ref{alg:reducing_complexity} further reduces $20\%$ vertices on average. However, the execution time would be 2-3 times slower than not applying it. Fig.~\ref{fig:exp_mesh_simplification} shows the remeshing results of the Egea model with and without applying initial mesh simplification, and Tab.~\ref{table:comparison} further compares the differences (OUR vs. OUR*). Usually, if users care more about the execution time than the mesh complexity, Alg.~\ref{alg:reducing_complexity} can be disabled. However, we enable Alg.~\ref{alg:reducing_complexity} by default for a better compromise among mesh complexity, element quality and approximation error.

\subsection{Evaluation of Final Vertex Relocation}
\label{sec:exp_post_subalgorithm}
\begin{figure}[hbt]
\centering
\includegraphics[width= 1.0\linewidth]{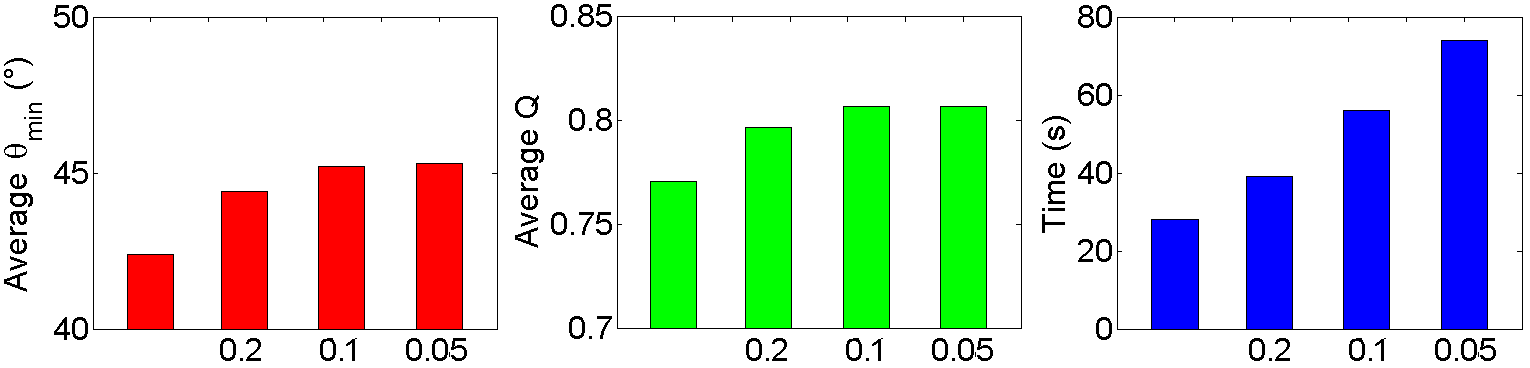} % \TODO{(This figure will be updated after new experiments by Kaimo)}
\caption{The effectiveness of $\Delta\theta$ in Alg.~\ref{alg:improving_quality}. In each sub figure, the first bar indicates the value when no final vertex relocation is applied. In this experiment, $\delta=0.2(\%bb)$ and $\theta=30^{\circ}$. The above data are the averages of 10 consecutive executions with the Helmet model (Table.~\ref{table:comparison}) as input.}
\label{fig:exp_post_relocate_quality}
\end{figure}

To measure how the overall element quality is improved by applying Alg.~\ref{alg:improving_quality}, we introduce two new measurements: the first is the average minimal angles of all triangles in $M_R$ and the second is the average value of triangle qualities defined as $Q_t=2\sqrt{3}S_t/(p_th_t)$~\cite{Frey97}, where $S_t$ is the area of triangle $t$, $p_t$ the in-radius of $t$ and $h_t$ the length of the longest edge in $t$. We tested the statistic element quality and the execution time with varying $\Delta\theta$ in Alg.~\ref{alg:improving_quality} (Fig.~\ref{fig:exp_post_relocate_quality}),
%Note that when the relocate count is set to 0, we mean Alg.~\ref{alg:improving_quality} is not applied. The results show that with the increase of the relocate count, the statistic {\color{revised}element quality} and the execution time increase accordingly. However,
and found when $\Delta\theta<0.1^{\circ}$, the quality improvement is not significant. In our experiments, we set the default $\Delta\theta$ as $0.1^{\circ}$ in Alg.~\ref{alg:improving_quality}.
%and enable Alg.~\ref{alg:improving_quality} by default.

\subsection{Influence of the Minimal Angle Threshold and the Mesh Complexity}
\label{sec:exp_minimal_angle}
Our algorithm produces results with either desired minimal angle threshold or desired mesh complexity, depending on the user specified parameters. In order to get the desired minimal angle, $N$ in Alg.~\ref{alg:remeshing} should be set very large; otherwise, $\theta$ should be set very large (e.g. $60^{\circ}$). Usually, the larger the value $\theta$ or $N$ is specified, the better element quality is achieved. We tested the remeshing results of the Fandisk model with $\theta$ varying from $0^{\circ}$ to $40^{\circ}$ and $N$ varying from 0.15k to 2.8k, and show the results in Fig.~\ref{fig:varying_theta}. The complete attributes are listed in Tab.~\ref{tab:varing_theta}. 
Note that in Fig. 1(g), one vertex has been relocated a little away from the crease after refinement, such that the minimal angle is improved. This happens when optimizing a mesh with sharp features up to a high minimal angle threshold.
%We find that when $\theta \leq 35^{\circ}$, the mesh complexity and execution time increases smoothly. However, when $\theta$ is up to $40^{\circ}$, the mesh complexity and execution time increases dramatically. 
%The RMS values decrease with the increase of $\theta$, since more vertex position optimizations are applied.
\begin{table}[hbt]
	\caption{Influence of $\theta$ and $\sigma$. $V_{567}$ is the percent of vertices with valences 5, 6 and 7. We set $\delta=0.2(\%bb)$, and the Fandisk (Fig.~\ref{fig:varying_theta}(a)) is the input.}
    \label{tab:varing_theta}
    \centering
    \begin{threeparttable}
    \begin{tabular}{|c|c|c|c|c|c|c|}
        \hline
$\bm{\theta}$ & \textbf{\#V} & $\bm{Q_{min}}$ & $\bm{\theta_{max}}$ & \textbf{RMS(\%bb)} & $\bm{V_{567}(\%)}$ & \textbf{Time}\\
		\hline
\textit{0}	& \textit{0.15k}	& 0.040		& 174.4		& 0.043	& 76.5		& 1:42*\\
\textit{10}	& \textit{0.20k}	& 0.188		& 155.1		& 0.038	& 71.3		& 2:18\\
\textit{20}	& \textit{0.26k}	& 0.340		& 135.5		& 0.036	& 78.6		& 2:41\\
\textit{30}	& \textit{0.40k}	& 0.482		& 117.7		& 0.031	& 81.7		& 3:01\\
\textit{35}	& \textit{0.73k}	& 0.552		& 109.3		& 0.029	& 85.5		& 3:44\\
\textit{40}	& \textit{2.8k}		& 0.640		& 98.8		& 0.022	& 98.5		& 6:14\\
        \hline
    \end{tabular}
    \begin{tablenotes}
	\item[*] This indicates the execution time of Alg.~\ref{alg:reducing_complexity}.
	\end{tablenotes}
	\end{threeparttable}
\end{table}

We compare our results with the state-of-the-art methods, and find only in a small portion of results presented in~\cite{Yan14A} and~\cite{Yan13}, the minimal angles exceed $35^{\circ}$ (with highest record $38^{\circ}$). For most other methods, the minimal angles vary between [$25^{\circ}$, $35^{\circ}$]. Contrary to the previous work, our method is able to generate results with minimal angles higher than $35^{\circ}$ in all test cases. The complete comparison with the state-of-the-art methods is shown in Sec.~\ref{sec:exp_comparison}.

\subsection{Influence of the Approximation Error Threshold}
\label{sec:exp_approximate_error}
\begin{figure}[hbt]
    \centering
    \begin{minipage}{44mm}
        \centering
        \includegraphics[width=44mm]{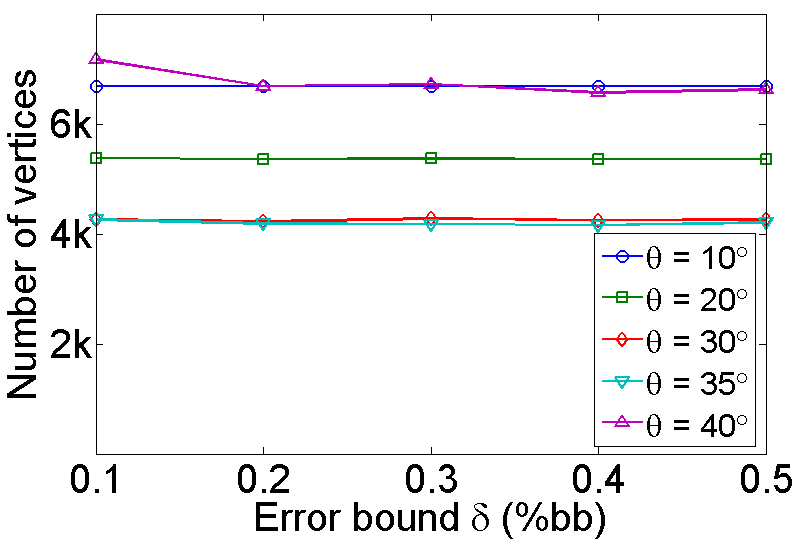}\\
    %{\small a. To be added.}
    \end{minipage}
    %\hspace{0.1in}
    \begin{minipage}{44mm}
        \centering
        \includegraphics[width=44mm]{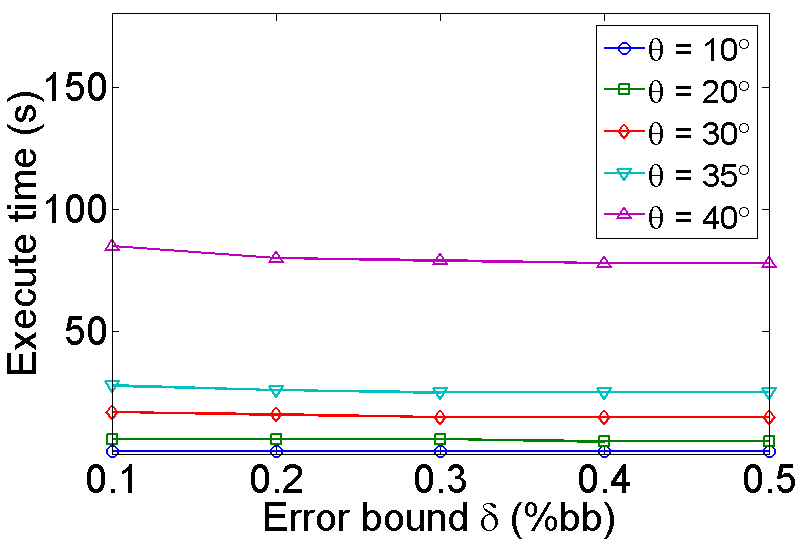}\\
    %{\small b. To be added.}
    \end{minipage}\\
    %\vskip0.1in
    \caption{Influence of the error-bound threshold $\delta$. The above data are the averages of 10 consecutive executions with the Elephant model (Table~\ref{table:comparison}) as input.}
\label{fig:exp_delta_variation}
\end{figure}

%The error-bound threshold $\delta$ is another important parameter for specifying the maximum error that the remeshing results are permitted to deviate from the input models. In theory, the higher value $\delta$ is specified, the more freedom the local operators can be selected to optimize the minimal angle.
%To investigate the influence of $\delta$ on the remeshing results, we tested the mesh complexity and execution time with different $\theta$ values, by setting varying $\delta$ values, as shown in Fig.~\ref{fig:exp_delta_variation}.
We demonstrate the influence of $\delta$ in Fig.~\ref{fig:exp_delta_variation}.
The results indicate that $\delta$ does not influence the mesh complexity and the execution time significantly with a fixed $\theta$ value. However, two interesting phenomena are observed: 1) when $\theta \leq 35^{\circ}$, the larger the $\theta$ value is, the lower the mesh complexity we achieve; however, when $\theta$ is set to $40^{\circ}$, both mesh complexity and execution time increase dramatically. This is because when $\theta$ is small, the edge collapse operator is preferentially applied, which increases the minimal angles while reduces the mesh complexity. However, when $\theta$ is large, more edge split operators are applied to modify local connections. 2) Within a fixed $\theta$, when $\delta$ increases, the mesh complexity decreases slightly and smoothly, since the higher $\delta$ is set, the more edge collapse operators are triggered.
%3) When $\delta = 0.5$, the mesh complexities of results in which $\theta \geq 35^{\circ}$ are a litter higher than the results in which $\theta < 30^{\circ}$. This is because when $\delta$ is high, the minimal angle is so easy and unconstrained to be optimized that less overall operators are required, making the final mesh complexities a little higher.

We compare our results with those provided by Yan et al.~\cite{Yan13, Yan14A, Yan14B}. The best record of the approximation error is $0.10(\%bb)$, which is generated by~\cite{Yan09} with the Homer model as input. However, it can not be explicitly controlled. For other methods, the approximation error is between $0.3-0.5(\%bb)$, and still cannot be strictly bounded. In striking contrast, our algorithm is able to explicitly and strictly control the approximation error, and achieve the approximation error as low as $0.07(\%bb)$ with the same input. More complete comparisons are in Sec.~\ref{sec:exp_comparison}.

\subsection{Robustness}
\label{sec:exp_robustness}
%To verify the robustness of our method, we apply it on a set of representative models. The results are shown in Fig.~\ref{fig:exp_results}.

\begin{figure}[hbt]
    \centering
    \begin{minipage}{28mm}
        \centering
        \includegraphics[width=28mm]{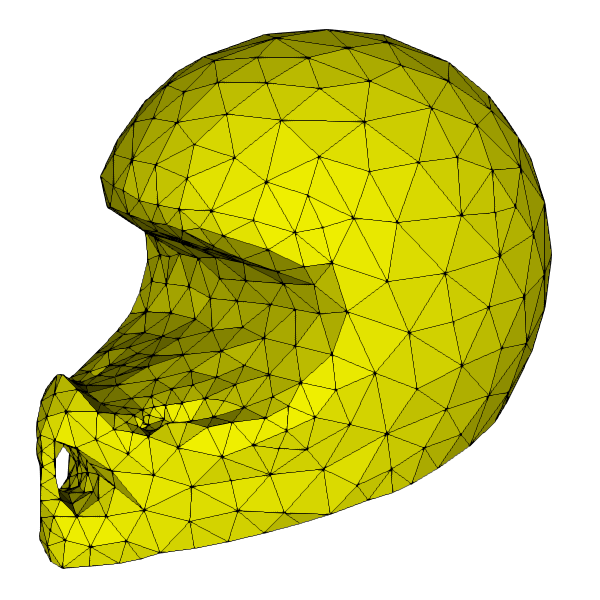}\\
    {\small (a) Helmet}
    \end{minipage}
    \begin{minipage}{28mm}
        \centering
        \includegraphics[width=28mm]{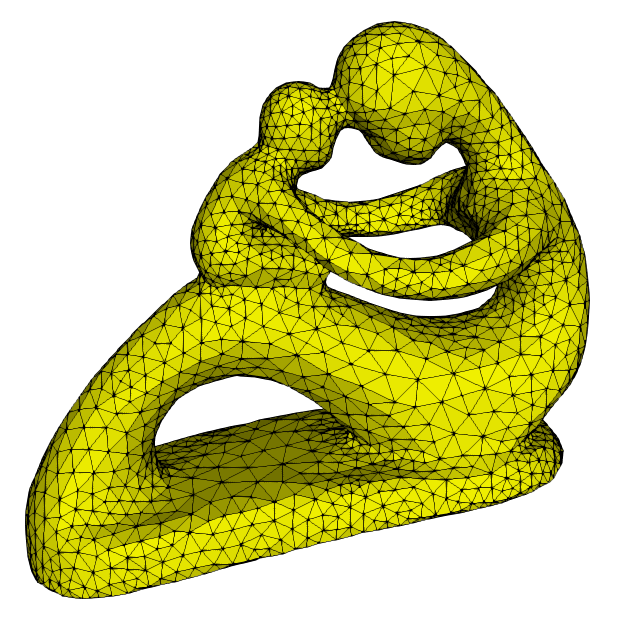}\\
    {\small (b) Fertility}
    \end{minipage}
    \begin{minipage}{28mm}
        \centering
        \includegraphics[width=28mm]{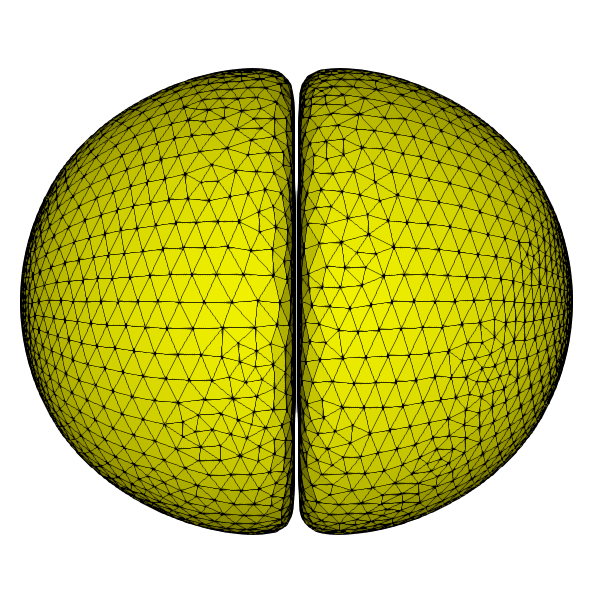}\\
    {\small (c) Close spheres}
    \end{minipage}\\
    %\vskip0.1in
    \begin{minipage}{28mm}
        \centering
        \includegraphics[width=28mm]{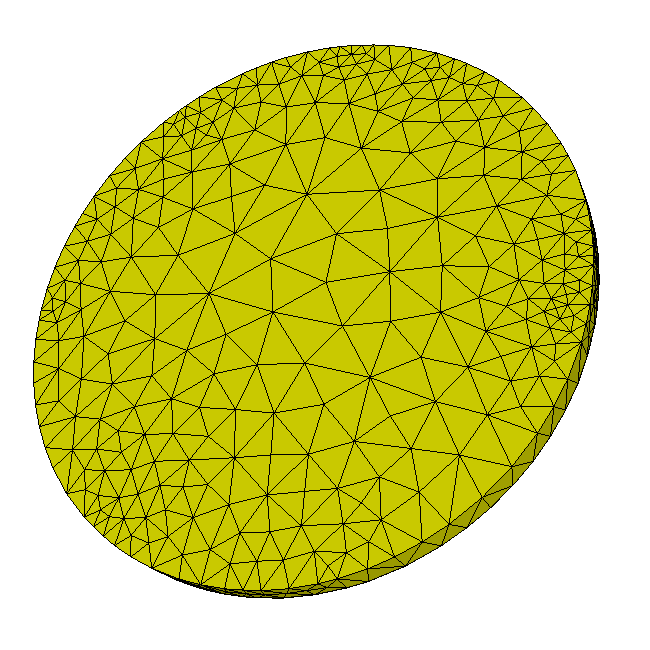}\\
    {\small (d) Disk}
    \end{minipage}
    \begin{minipage}{28mm}
        \centering
        \includegraphics[width=28mm]{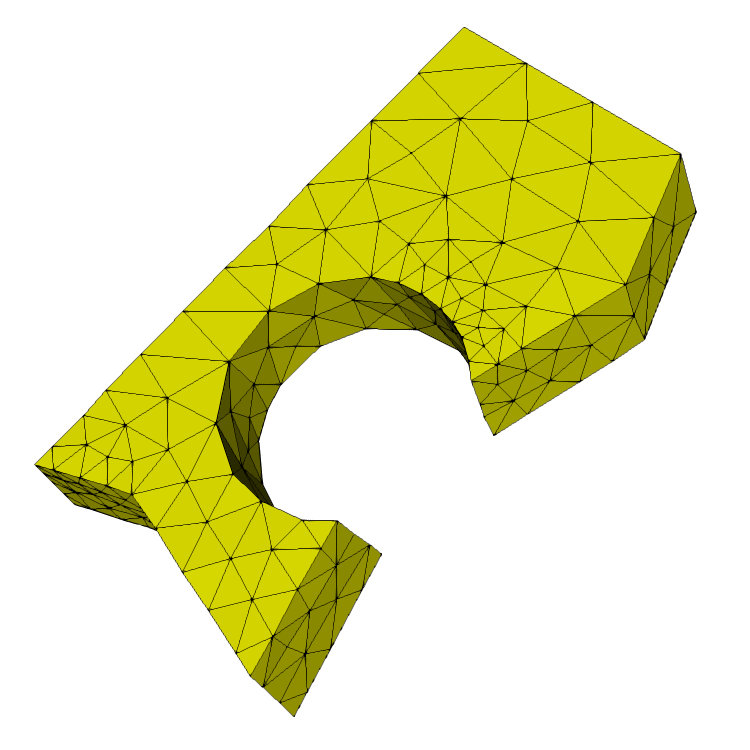}\\
    {\small (e) U-part}
    \end{minipage}
    \begin{minipage}{28mm}
        \centering
        \includegraphics[width=28mm]{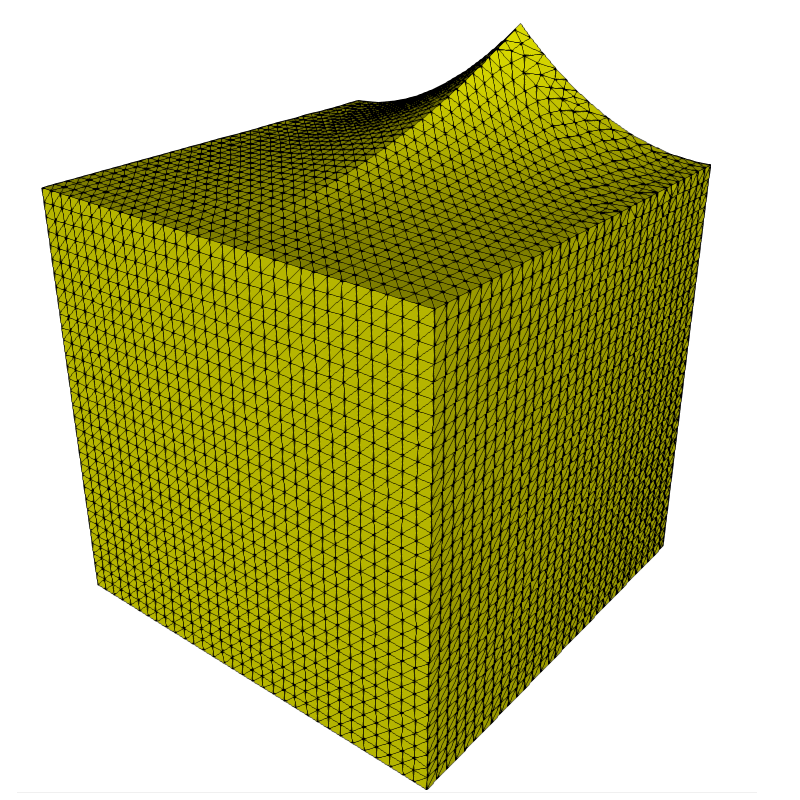}\\
    {\small (f) Smooth crease}
    \end{minipage}
    %\vskip0.1in
    \begin{minipage}{28mm}
        \centering
        \includegraphics[width=28mm]{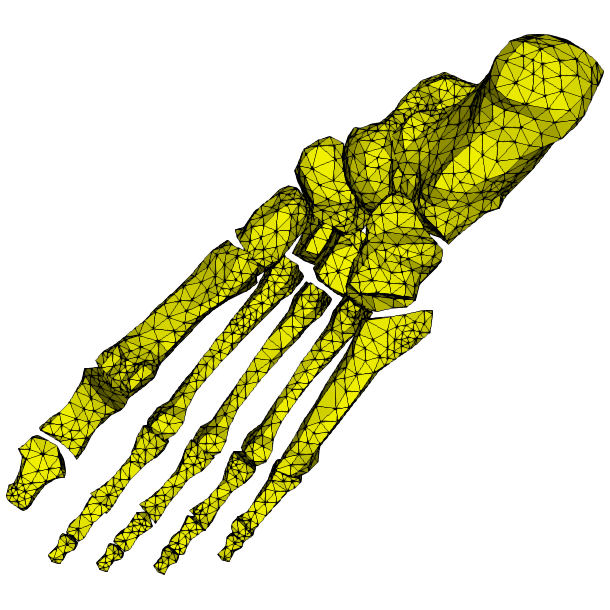}\\
    {\small (g) Bones}
    \end{minipage}
    \begin{minipage}{28mm}
        \centering
        \includegraphics[width=28mm]{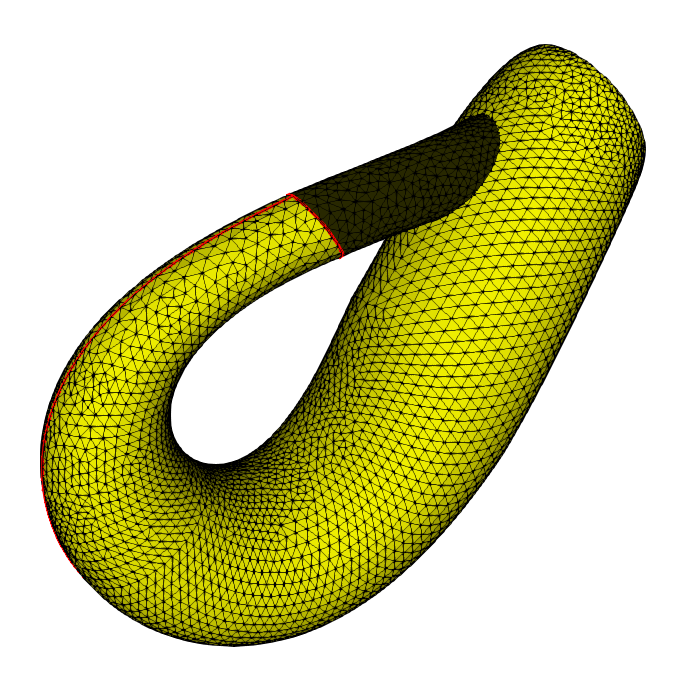}\\
    {\small (h) Klein's bottle}
    \end{minipage}
    \begin{minipage}{28mm}
        \centering
        \includegraphics[width=28mm]{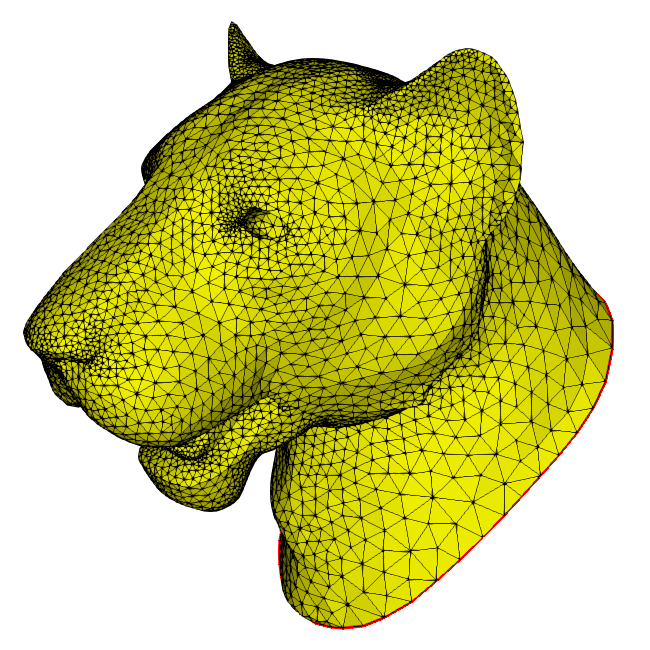}\\
    {\small (i) Lion head}
    \end{minipage}
    \caption{Selected results. In this experiment, $\delta$ is set to $0.2\%(bb)$ and $\theta$ is set to $35^{\circ}$. In (h) and (i), the boundaries are rendered in red. The dark part of (h) means the triangle normals are inside.}
\label{fig:exp_results}
\end{figure}

Since in our local error update scheme, the closest point pairs of stratified samples are reliably initialized
%(the closest point of a sample is initialized as itself)
and locally updated, our method is robust to models with complex topology, holes, and intersections of surfaces. For example, the Close spheres model (Fig.~\ref{fig:exp_results}(c)) is composed of two rounded half spheres that are positioned very close to each other, and the Klein bottle model (Fig.~\ref{fig:exp_results}(h)) exhibits non-orientable surfaces that are self-intersecting. Our method generates the right results for both of them.

By integrating the feature intensity function, our method successfully handles models with and without clear sharp features. The Helmet (Fig.~\ref{fig:exp_results}(a)), Fertility (Fig.~\ref{fig:exp_results}(b)) and Close spheres models are smooth, whereas the U-part (Fig.~\ref{fig:exp_results}(e)) and the Smooth crease models (Fig.~\ref{fig:exp_results}(f)) have sharp features.
%Our method generates satisfying results for all of them without explicit feature detection or specification. 
In addition, since the feature intensity also captures boundaries, our method is capable of remeshing models with boundaries (Fig.~\ref{fig:exp_results}(i)) and smooth features (Fig.~\ref{fig:exp_results}(f)).

Our method requires no surface parameterization, making it naturally suitable for high-genus models (Fig.~\ref{fig:exp_results} (a) and Fig.~\ref{fig:exp_results}(b)) as well as models with multiple components (Fig.~\ref{fig:exp_results}(g)). Note that adaptivity is created automatically by the approximation error parameter $\delta$ if reasonable without requiring an a priori estimation of a density function. 
%Given an input model, fewer vertices will be generated in flat areas, whereas more vertices are automatically generated in high curvature areas.
%This flexibility is solely controlled by the parameter $\delta$. 
The Helmet, Fertility, U-part and Lion head models in Fig.~\ref{fig:exp_results} clearly illustrate this advantage.

%Even though the mesh complexity of the results is highly related to the input models
Our method is suitable to process models with very high/low resolutions and/or badly shaped triangles. To tackle dense models, typically the initial mesh simplification strongly reduces the mesh complexity. For very coarse models, users can optionally increase the sampling density for better error control. We present two typical examples: the input Fertility model has 13k vertices, whereas our remeshing result (Fig.~\ref{fig:exp_results}(b)) has only 2.9k vertices, thanks to the effectiveness of Alg.~\ref{alg:reducing_complexity}. The input U-part model possesses only 86 vertices. By sampling 50 points in each facet averagely, we get the result with 347 vertices, and the geometric fidelity is well-controlled.

\subsection{Comparisons}
\label{sec:exp_comparison}
We compare our approach to the state-of-the-art techniques in terms of efficiency, geometric fidelity, element quality and mesh complexity. For simplicity, only the most efficient methods (RAR~\cite{Dunyach13} and MMGS, an improvement of YAMS~\cite{Frey00}) and the methods that produce the best results in Yan and Wonka's conduction~\cite{Yan13} (CVT~\cite{Yan09} (100 iterations in our experiments), CVD~\cite{Valette08} and MPS~\cite{Yan13}) are compared with identical inputs.
%Since their comparisons were performed on an Intel C5680 Dual Core 3.33GHz CPU with 4GB memory, the timings are not directly comparable, but can be regarded as a reference.
Among all the compared methods, CVT and CVD require the number of vertices to be specified, and the Hdist is required in RAR and MMGS. To make the results comparable, we set the Hdist of RAR to the same value as that of our method, and carefully adjusted the Hdist parameter for MMGS, such that it generates results with the same complexity as CVT, MPS and CVD. In our method, OUR* means Alg.~\ref{alg:reducing_complexity} is disabled while OUR means it is enabled. A detailed comparison is listed in Tab.~\ref{table:comparison}, and Fig.~\ref{fig:exp_compare} shows a close-up comparison. More visual comparisons are provided in the supplemental materials, available online.

\begin{table*}
    \caption{Comparison with the state-of-the-art methods. For all the method, the input parameters are highlighted with italic fonts, and the best results are highlighted in bold. In column $Hdist (\%bb)$, the values before ``$/$'' are the input, and the values after ``$/$'' are the real HDist measured by Metro~\cite{cignoni98}.}
    \label{table:comparison}
    \centering
    \begin{tabular}{c|cccccccccc}
        \hline
        \hline
        \textbf{Input}	& \textbf{Methods}	& $\bm{\#V}$ & $\bm{Q_{min}}$ & $\bm{\theta_{min} (^\circ)}$ & $\bm{\theta_{max} (^\circ)}$ & \textbf{Hdist (\%bb)} & \textbf{RMS (\%bb)} & $\bm{\theta < 30^\circ(\%)}$ & $\bm{V_{567}(\%)}$ & \textbf{Time}\\
        \hline
Rockerarm	& [RAR]	& \textbf{2.1k}	& 0.556	& 27.9	& 107.2	& \textit{0.20}/0.94	& 0.120	& 0.02	& \textbf{100}	& $\bm{<0:01}$\\
(3.4k)		& [MMGS]& 5.8k	& 0.056	& 3.4	& 172.5	& \textit{0.47}		& 0.103	& 1.73	& 93.4	& $0:01$\\
			& [CVT]	& \textit{5.8k}	& 0.588	& 28.3	& 104.6	& 0.21		& 0.030	& 0.02	& 99.9	& $0:48$\\
			& [MPS]& \textit{5.8k}	& 0.516	& 32.0	& 113.6	& 0.48		& 0.033	& \textbf{0}		& \textbf{100}	& $0:05$\\
			& [OUR*] & 2.8k & 0.559	& \textit{35.0}	& 108.5	& \textit{0.20}/0.20	& 0.025	& \textbf{0}		 & 89.0	& $0:29$\\
			& [OUR*] & \textit{3.0k} & 0.612	& 38.6	& 102.2	& \textit{0.20}/0.17	& 0.024	& \textbf{0}		 & 95.4	& $0:38$\\
			& [OUR*]& 3.8k	& \textbf{0.646}	& \textbf{\textit{40.0}}	& \textbf{98.1}	& \textit{0.20}/\textbf{0.14}	 & \textbf{0.020}	& \textbf{0}		& 98.5	& $0:51$\\
			& [OUR]	& 3.6k	& 0.639	& \textbf{\textit{40.0}}	& 99.0	& \textit{0.20}/0.20	& 0.025	& \textbf{0}		 & 97.9	& $2:58$\\
        \hline
Homer		& [RAR]	& \textbf{2.6k}	& 0.569	& 29.2	& 106.5	& \textit{0.20}/0.55	& 0.070	& 0.02	& 99.9	 & $\bm{<0:01}$\\
(6.0k)		& [MMGS]& 7.2k	& 0.210	& 13.1	& 152.2	& \textit{0.43}		& 0.028	& 1.07	& 95.8	& $0:01$\\
			& [CVT]	& \textit{7.2k}	& 0.568	& 25.3	& 102.3	& 0.10		& 0.021	& 0.02	& 99.9	& $1:14$\\
			& [MPS]& \textit{7.2k}	& 0.513	& 32.0	& 115.0	& 0.31		& 0.023	& \textbf{0}		& \textbf{100}	& $0:05$\\
			& [OUR*]& 4.8k	& 0.553	& \textit{35.0}	& 109.2	& \textit{0.20}/0.09	& 0.010	& \textbf{0}		 & 91.8	& $0:46$\\
			& [OUR*] & \textit{5.0k} & 0.600	& 37.8	& 103.6	& \textit{0.20}/0.09	& 0.010	& \textbf{0}		 & 95.2	& $0:59$\\
			& [OUR*]& 6.9k	& \textbf{0.643}	& \textbf{\textit{40.0}}	& \textbf{98.5}	& \textit{0.20}/\textbf{0.07}	 & \textbf{0.009}	& \textbf{0}		& 98.7	& $1:29$\\
			& [OUR]	& 4.3k	& 0.635	& \textbf{\textit{40.0}}	& 99.5	& \textit{0.20}/0.17	& 0.018	& \textbf{0}		 & 97.8	& $4:48$\\
        \hline
Triceratops	& [RAR]	& \textbf{1.6k}	& 0.607	& 30.0	& 98.4	& \textit{0.20}/2.61	& 0.570	& 0.03	& 99.8	 & $\bm{<0:01}$\\
(2.8k)		& [MMGS]& 9.0k	& 0.270	& 13.5	& 143.3	& \textit{0.41}		& 0.080	& 1.11	& 93.6	& $0:01$\\
			& [CVT]	& \textit{9.0k}	& 0.543	& 31.7	& 110.3	& \textbf{0.12}		& \textbf{0.018}	& \textbf{0}	& 99.9	& $1:23$\\
			& [MPS]& \textit{9.0k}	& 0.506	& 32.0	& 114.8	& 0.46		& 0.062	& \textbf{0}		& \textbf{100}	& $0:29$\\
			& [OUR*]& 2.1k	& 0.552	& \textit{35.0}	& 109.3	& \textit{0.20}/0.16	& 0.028	& \textbf{0}		& 87.0	& $0:28$\\
			& [OUR*] & \textit{3.0k} & 0.605	& 38.4	& 103.0	& \textit{0.20}/0.18	& 0.024	& \textbf{0}		 & 93.8	& $0:41$\\
			& [OUR*]& 4.8k	& 0.634	& \textbf{\textit{40.0}}	& 99.6	& \textit{0.20}/0.19	& 0.036	& \textbf{0}		 & 97.7	& $1:39$\\
			& [OUR]	& 3.5k	& \textbf{0.644}	& \textbf{\textit{40.0}}	& \textbf{98.4}	& \textit{0.20}/0.19	& 0.040	& \textbf{0}		& 97.1	& $2:36$\\
        \hline
Elephant	& [RAR]	& 2.9k	& 0.480	& 24.9	& 115.5	& \textit{0.20}/5.0	& 0.112	& 0.12	& \textbf{100}	& $\bm{<0:01}$\\
(6.9k)		& [MMGS]& 11k	& 0.187	& 10.9	& 155.1	& \textit{0.29}		& 0.034	& 1.24	& 93.5	& $0:02$\\
        	& [CVT]	& \textit{11k}	& 0.560	& 26.8	& 107.7	& \textbf{0.11}		& 0.018	& 0.01	& 99.8	& $1:43$\\
        	& [MPS]& \textit{11k}	& 0.505	& 32.0	& 114.9	& 0.38		& 0.061	& \textbf{0}		& \textbf{100}	& $0:31$\\
        	& [OUR*]& 4.4k	& 0.553	& \textit{35.0}	& 109.2	& \textit{0.20}/\textbf{0.11}	& 0.014	& \textbf{0}		& 90.4	& $0:49$\\
        	& [OUR*] & \textit{5.0k} & 0.617	& 39.1	& 101.5	& \textit{0.20}/0.13	& 0.013	& \textbf{0}		 & 96.6	& $1:05$\\
        	& [OUR*]& 6.8k	& \textbf{0.638}	& \textbf{\textit{40.0}}	& \textbf{99.1}	& \textit{0.20}/\textbf{0.11}	 & \textbf{0.010}	& \textbf{0}		& 98.7	& $1:45$\\
        	& [OUR]	& \textbf{2.7k}	& 0.633	& \textbf{\textit{40.0}}	& 99.7	& \textit{0.20}/0.14	& 0.021	& \textbf{0}		& 98.4	& $4:25$\\
        \hline
Bunny		& [RAR]	& 4.1k	& 0.545	& 27.5	& 109.4	& \textit{0.20}/0.69	& 0.072	& 0.02	& \textbf{100}	 & $\bm{<0:01}$\\
(34k)		& [MMGS]& 12k	& 0.260	& 13.1	& 142.4	& \textit{0.41}		& 0.041	& 0.59	& 93.8	& $0:03$\\
        	& [CVD]	& \textit{12k}	& 0.150	& 9.6	& 160.1	& 0.34		& 0.028	& 0.71	& 96	& $0:05$\\
            & [CVT]	& \textit{12k}	& 0.603	& 30.6	& 103.1	& 0.20		& 0.018	& \textbf{0}	& 99.9	& $3:57$\\
            & [MPS]& \textit{12k}	& 0.510	& 32.0	& 114.6	& 0.37		& 0.035	& \textbf{0}		& \textbf{100}	& $0:24$\\
            & [OUR*]& 37k	& 0.550	& \textit{35.0}	& 109.6	& \textit{0.20}/\textbf{0.07}	& \textbf{0.003}	& \textbf{0}		& 95.8	& $3:23$\\
            & [OUR] & \textbf{\textit{1.5k}} & 0.637	& 39.8	& 99.2	& \textit{0.20}/0.19	& 0.033	& \textbf{0}		 & 97.2	& $10:41$\\
            & [OUR]	& 1.7k	& \textbf{0.646}	& \textbf{\textit{40.0}}	& \textbf{98.2}	& \textit{0.20}/0.19	 & 0.031	& \textbf{0}		& 97.7	& $12:28$\\
        \hline
        \hline
    \end{tabular}
\end{table*}

From all compared methods, RAR performs most efficiently and introduces the lowest mesh complexity. However, the geometric fidelity cannot be guaranteed.
%since the target edge lengths are estimated using local curvatures, which are not exact when the inputs are coarse. 
MMGS is also efficient, yet it introduces much higher Hdist distances. Our method is almost at the same level of efficiency as CVT when $\theta=35^{\circ}$, but is much slower when $\theta=40^{\circ}$.

\begin{figure*}[!hbt]
    \centering
    \begin{minipage}{50mm}
        \centering
        \includegraphics[height=50mm]{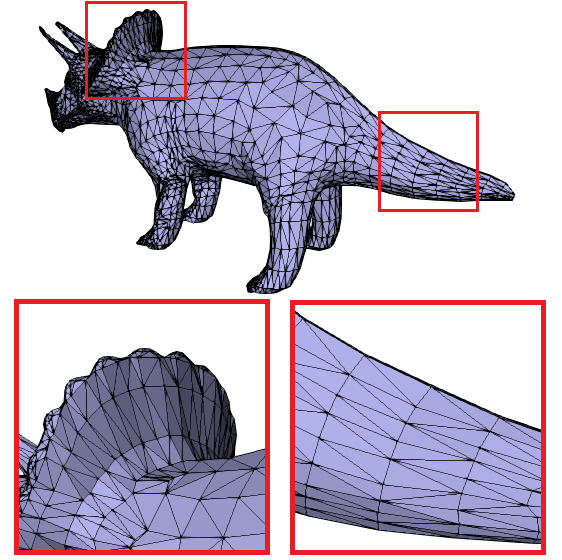}\\
    {\small (a) Input Triceratops model}
    \end{minipage}
    \begin{minipage}{25mm}
        \centering
        \includegraphics[height=50mm]{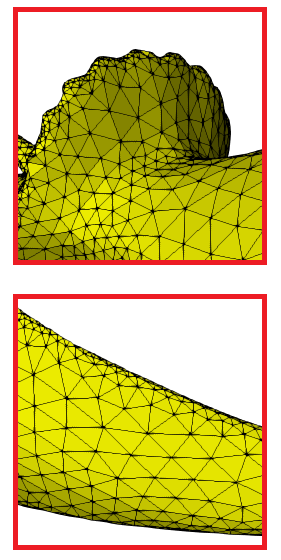}\\
    {\small (b) RAR}
    \end{minipage}
    \begin{minipage}{25mm}
        \centering
        \includegraphics[height=50mm]{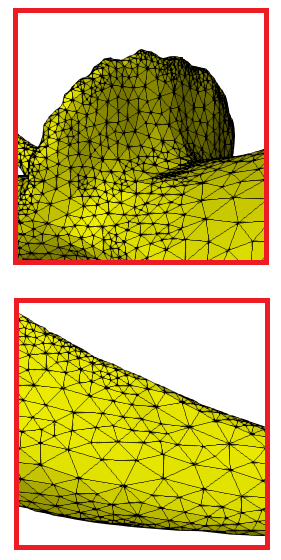}\\
    {\small (c) MMGS}
    \end{minipage}
    \begin{minipage}{25mm}
        \centering
        \includegraphics[height=50mm]{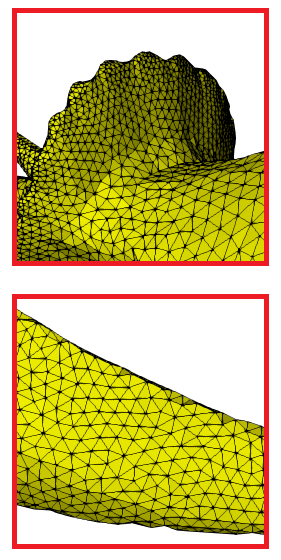}\\
    {\small (d) CVT}
    \end{minipage}
    \begin{minipage}{25mm}
        \centering
        \includegraphics[height=50mm]{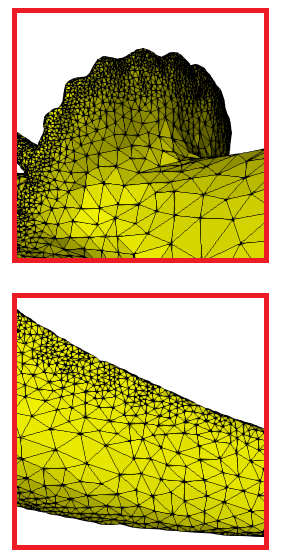}\\
    {\small (e) MPS}
    \end{minipage}
    \begin{minipage}{25mm}
        \centering
        \includegraphics[height=50mm]{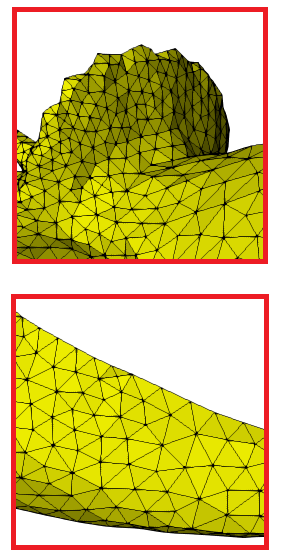}\\
    {\small (f) OUR}
    \end{minipage}\\
    %\vskip0.1in
    \caption{A close-up comparison of results with state-of-the-art approaches. In our method, $\delta$ is set to $0.20(\%bb)$ and $\theta$ is set to $40^{\circ}$.}
\label{fig:exp_compare}
\end{figure*}

According to~\cite{Yan13}, CVT performs best in keeping high geometric fidelity, but cannot explicitly bound the approximation error. By setting $\delta$ comparable to CVT's best results, our method consistently produces results with strictly bounded Hdist, and produces lower Hdist and RMS distances in most cases. 
%This is because our method modifies the input models as little as possible when eliminating the minimal angle, and moreover, the vertex position optimization procedure further reduces the approximation error.
%Among existing methods, MPS generates results with the highest minimal angle, best triangle quality, and regularity. 
We also find that our method consistently generates results with higher minimal angle and triangle quality than MPS, due to the fact that our method provides a better improvement of ``worst element quality'' measured by the minimal angle. However, since our method does not optimize the global connectivity of the input, our results have lower regularity (measured as $V_{567}$) than MPS.

%In our algorithm, the mesh complexity can be controlled explicitly as well. 
From Tab.~\ref{table:comparison}, we see that by setting the desired resolution lower than MMGS, CVT, MPS and CVD, we still get results with higher geometric fidelity and better bounds of minimal angle. For the very dense Bunny model, when Alg.~\ref{alg:reducing_complexity} is enabled, our method even reduces the complexity of the Bunny model to $5\%$ of the input without violating the error-bound constraint. When it is disabled, the resolutions are still lower than the inputs in most cases, since the edge collapse operator has high priority in Alg.~\ref{alg:remeshing}.

%Another advantage of our method is the automatic feature preservation (Fig.~\ref{fig:varying_theta} and Fig.~\ref{fig:exp_results}). 
To the best of our knowledge, RAR, MMGS, CVT and MPS require sharp features to be specified or detected in advance, which may be time-consuming or error-prone. Though CVD is able to preserve features implicitly, it leads to results with lower geometric fidelity and element quality than our method.
%On the contrary, our method preserves features implicitly yet produces results with better geometric fidelity and lower mesh complexity.

Since our method does not optimize the element quality globally, the average element quality is not superior to remeshing based on farthest point optimization~\cite{Yan14B} (FPO).
%which produces the best average {\color{revised}element quality} in literature.
%Though the subsequent vertices relocate subalgorithm can be optionally applied in our algorithm to further improve the average element quality, our results are still not as good as.
However, we consistently produce results with better geometric fidelity and larger minimal angle than FPO.

\subsection{Limitations}
\label{sec:exp_limitation}
Although practically the minimal angles can be improved to values above  $35^{\circ}$ in all our test cases, we do not have any theoretical guarantee for the convergence with a specified $\theta$. To challenge our algorithm, we set $\theta$ to its theoretical upper bound and show the best results that our algorithm achieved in Tab.~\ref{tab:fail_case}. We found the approximation error can still be bounded. However, the algorithm runs into infinite loops or generates degenerated edges while refining.

Another limitation is that we can only tackle 2-manifold meshes, for the reason that our local operators highly rely on the topology information of local regions. Finally, our method cannot remesh noisy models robustly, since the feature intensity function will interpret the noise as some kind of features and thus tries to preserve it.

\begin{table}[hbt]
	\caption{Cases when too high $\theta$ is specified. We set $\theta = 60^{\circ}$, run Alg.~\ref{alg:remeshing} until it fails, and record the best results it achieved. Alg.~\ref{alg:reducing_complexity} is enabled here.}
    \label{tab:fail_case}
    \centering
    \begin{tabular}{|c|c|c|c|c|c|c|}
        \hline
\textbf{Input} & \textbf{\#V} & $\bm{\theta_{min}}$ & \textbf{Hdist(\%bb)} & \textbf{Fail Types}\\
		\hline
Rockerarm	& 4.2k	& 41.22	& \textit{0.20}/0.19	& Degenerated edges\\
Homer		& 6.1k	& 41.13	& \textit{0.20}/0.14	& Infinite loops\\
Triceratops	& 4.8k	& 40.02 & \textit{0.20}/0.19	& Degenerated edges\\
Elephant	& 4.8k	& 41.23	& \textit{0.20}/0.13	& Infinite loops\\
Bunny		& 2.4k	& 41.42	& \textit{0.20}/0.19	& Infinite loops\\
        \hline
    \end{tabular}
\end{table}

\section{Conclusion}
\label{sec:conclusion}
We presented a novel surface remeshing algorithm based on minimal angle improvement. In this framework, the minimal angle of the input model is sequentially improved by applying local operators. Furthermore, an efficient and reliable local error update scheme was designed and embedded for explicitly bounding the approximation error, and two novel feature intensity functions were defined and integrated in the vertex relocation, in order to preserve features implicitly. Compared to the state-of-the-art, our method consistently generates results with higher element quality and lower mesh complexity but satisfied error-bounds. The resulting meshes are well suited for robust numerical simulations, since they offer bounded approximation error and large minimal angles.

However, there are still some limitations (Sec.~\ref{sec:exp_limitation}), which motivate future work in the following aspects: 1) providing a theoretical convergence guarantee with a specified $\theta$; 2) extending the current implementation such that this framework can be applied to triangle soups or even point clouds, for error-bounded and feature preserving mesh reconstruction; and 3) exploring more robust feature intensity functions for measuring sharp features on noisy models.

% or


%\appendices

%\section{}

\ifCLASSOPTIONcompsoc
  \section*{Acknowledgments}
\else
  \section*{Acknowledgment}
\fi
We wish to thank Mario Botsch for providing the source code and results of their method, and Pascal Frey for providing the MMGS software. This work was partially supported by the European Research Council (ERC Starting Grant Robust Geometry Processing \#257474), the National Science Foundation of China (61373071, 61372168, 61620106003) and the German Research Foundation (DFG, grant GSC 111, Aachen Institute for Advanced Study in Computational Engineering Science).

\ifCLASSOPTIONcaptionsoff
  \newpage
\fi



\bibliographystyle{IEEEtran}
\bibliography{TVCG_remesh}
\end{document}